\documentclass[aip,pof,preprint]{revtex4-2}

\usepackage{graphicx}
\usepackage{subcaption} 

\makeatletter
\def\@email#1#2{%
	\endgroup
	\patchcmd{\titleblock@produce}
	{\frontmatter@RRAPformat}
	{\frontmatter@RRAPformat{\produce@RRAP{*#1\href{mailto:#2}{#2}}}\frontmatter@RRAPformat}
	{}{}
}%
\makeatother
\begin{document}
	\newcommand{\Nu}{\mathit{Nu}}
	\newcommand{\Ra}{\mathit{Ra}}
	\newcommand{\Rif}{\mathit{Ri}_f}
	\newcommand{\Pran}{\mathit{Pr}}
	\newcommand{\Rey}{\mathit{Re}}
	
	
	\title{Analysis and reformulation of the $k$--$\omega$ turbulence model for buoyancy-driven thermal convection}

	\author{Da-Sol Joo\footnote{Email address: \texttt{wnekthf3818@postech.ac.kr}}}
	\affiliation{Department of Mechanical Engineering, Pohang University of Science and Technology, 77 Cheongam-ro, Nam-gu, Pohang, Gyeongbuk 37673, Republic of Korea}

	\date{\today}
	
	\begin{abstract}
		The representation of buoyancy-driven turbulence in Reynolds-averaged Navier--Stokes (RANS) models remains unresolved, with no widely accepted standard formulation. 
		A key difficulty is the lack of analytical guidance for incorporating buoyant effects, particularly under unstable stratification. 
		This study derives an analytical solution of the standard $k$--$\omega$ model for Rayleigh--B\'enard convection in an infinite layer, where turbulent kinetic energy is generated solely by buoyancy. 
		The solution provides explicit scaling relations among the Rayleigh ($\Ra$), Prandtl ($\Pran$), and Nusselt ($\Nu$) numbers that capture the simulation trends: $\Nu \sim \Ra^{1/3}\Pran^{1/3}$ for $\Pran \ll 1$ and $\Nu \sim \Ra^{1/3}\Pran^{-0.415}$ for $\Pran \gg 1$.
		This framework quantifies the discrepancies in the conventional buoyancy treatment and clarifies their origin.
		Informed by this analysis, the buoyancy-related modeling terms are reformulated to recover the measured trends: namely $\Nu \sim \Pran^{1/8}$ for $\Pran \ll 1$ and $\Nu \sim \Pran^{0}$ for $\Pran \gg 1$ at moderate $\Ra$.
		Only two dimensionless algebraic functions are introduced, which vanish in the absence of buoyancy, ensuring full compatibility with the standard closure.
		The corrected model is validated across a range of buoyancy-driven flows, including two-dimensional Rayleigh--B\'enard convection, internally heated convection in two configurations, unstably stratified Couette flow, and vertically heated natural convection with varying aspect ratios. 
		Across all cases, the corrected model provides significantly improved predictions of mean temperature fields and turbulent heat flux distributions.
	\end{abstract}
	
	\maketitle
	
\section{Introduction}\label{sec:1}

Buoyancy-driven turbulence is ubiquitous in nature, spanning the oceans and atmosphere, and plays a central role in engineering applications such as indoor ventilation and nuclear safety. For such flows, Reynolds-averaged Navier--Stokes (RANS) closures remain widely used in engineering practice because of their favorable accuracy-to-cost ratio, with two-equation models---notably the standard $k$--$\varepsilon$ and $k$--$\omega$ formulations---being particularly prevalent.

Comprehensive references and review articles on buoyancy-related RANS modeling can be found in 
\citet{lumley1979computational,rodi1980turbulence,rodi1987examples,hanjalic2002one,launder2005rans,burchard2007applied,durbin2011statistical,hanjalic2011modelling,durbin2018some,hanjalic2021reassessment}. 
In natural convection, temperature acts as an active scalar that modifies the momentum field through buoyancy. The exact transport equation for turbulent kinetic energy ($k$) contains an explicit buoyant-production term proportional to the inner product between the turbulent heat flux and gravity; this term is negative in stable stratification and positive in unstable stratification. Because the modeled $k$ equation in both the $k$--$\varepsilon$ and $k$--$\omega$ frameworks is derived from this exact equation, it is standard practice to include buoyant production in $k$.

The central controversy concerns the scale-determining equation---namely, the dissipation ($\varepsilon$) or dissipation-per-unit-$k$ ($\omega$) equation---which is typically the sole source of buoyancy-related modeling terms in two-equation RANS closures. 
As summarized in reference works \citep{burchard2007applied,hanjalic2011modelling} and in several studies \citep{henkes1991natural,peng1999computation,hanjalic1993computation,hanjalic2021reassessment}, there is no universally accepted prescription for representing buoyancy at this level.

Broadly, two lines of work can be distinguished. 
One omits any buoyant contribution from the scale-determining equation. The rationale is that shear production in this equation can be justified based on canonical shear-flow analyses, whereas an analogous justification for a buoyant counterpart in $\varepsilon$ or $\omega$ is lacking \citep{markatos1984laminar}. 
The other line of work introduces a buoyancy term into the scale-determining equation, modeling it by analogy with shear production and adopting the same functional form as the shear-production term in the $\varepsilon$- or $\omega$-equation.

As described by \citet{hanjalic1993computation}, in flows such as Rayleigh--B\'enard convection---where shear production is nearly zero but buoyant production dominates---omitting the buoyant-production term from the $\varepsilon$ equation can render $\varepsilon$ a pure sink. When combined with finite buoyant production in $k$, this may lead to unphysical growth of $k$. 
A number of empirical blending strategies have also been proposed for specific regimes \citep{rodi1980turbulence,rodi1987examples,henkes1991natural,hanjalic1993computation,peng1999computation}. A frequently cited guideline is to suppress buoyant contributions under stable stratification while including them under unstable stratification, using the same functional form as the shear-production term \citep{rodi1987examples}. 
Each of these studies explicitly acknowledged the empirical nature of their proposed treatments, and the proper representation of buoyancy in two-equation RANS closures remains an open issue.

The absence of a unified criterion stems from the lack of systematic analyses establishing how standard two-equation closures behave in canonical, buoyancy-dominated flows. By contrast, in shear-driven canonical flows, the analytical behavior of standard RANS models is well established: they reproduce classical results such as the logarithmic velocity law with the von K\'arm\'an constant, boundary-layer scaling, spreading rates of free-shear flows, and self-similarity in homogeneous shear and decaying turbulence. These benchmarks have served both to justify the formulation of the models and to calibrate their coefficients \citep{speziale1990analytical,pope2000turbulent,wilcox2006turbulence}.

Specifically, to the best of the author's knowledge, analytical criteria are available for modeling stably stratified flows, whereas corresponding criteria for unstable stratification remain scarce. In the stable regime, linear-stability analyses of stably stratified shear flows provide a clear physical measure of flow suppression, commonly expressed in terms of the flux Richardson number. This quantity, defined locally as the ratio of buoyant to shear production, reaches a critical value of $\mathit{Ri}_f \approx 0.25$, beyond which stratification begins to stabilize the shear flow. Owing to this established threshold, $\mathit{Ri}_f$ has long served as a practical modeling reference \citep{gibson1978ground,durbin2011statistical}, guiding the selection of model constants in $k$--$\varepsilon$ closures \citep{burchard1995performance} and informing explicit algebraic closures for stable stratification \citep{lazeroms2013explicit}.

By contrast, in unstably stratified buoyant flows, an appropriate model should represent the rate at which potential energy associated with density variations is converted into the kinetic energy of thermal plumes---a mechanism fundamentally different from shear production. In homogeneous shear flow, the dynamics are primarily governed by the shear rate $S$, $k$, and $\varepsilon$, yielding the local similarity relation $\varepsilon/k \sim S$, which is widely accepted as a standard modeling benchmark \citep{pope2000turbulent,wilcox2006turbulence}. 
For buoyant effects, however, evidence of such purely local self-similarity remains scarce. This limitation is often cited as an indirect justification for omitting a buoyancy term from the scale-determining equation, as in \citet{markatos1984laminar}.

Instead, similarity in buoyant flows is typically expressed through global power-law relations among the Nusselt ($\Nu$), Rayleigh ($\Ra$), and Prandtl ($\Pran$) numbers---for example, the scaling $\Nu \sim \Ra^{0.31}$ observed over $10^{6}\lesssim \Ra \lesssim 10^{17}$ for $\Pran\approx 0.7$ by \citet{niemela2000turbulent}. Because these relations depend on global temperature differences and a domain-scale length, the assessment of buoyancy effects in two-equation closures is more appropriately undertaken through domain-integrated analyses rather than purely local arguments.

Against this background, the present study adopts statistically one-dimensional Rayleigh--B\'enard convection as a canonical test bed for analyzing buoyancy effects in the Wilcox (2006) $k$--$\omega$ model \citep{wilcox2006turbulence}. In this configuration, the mean flow vanishes and the turbulent kinetic energy budget is governed solely by buoyant production, thereby isolating the role of the buoyancy term in the $\omega$ equation. This case corresponds to Rayleigh--B\'enard convection in a horizontally periodic plane layer, for which direct numerical simulation (DNS) budgets clearly show buoyant production dominating the $k$ balance \citep{togni2015physical}.

The choice of the $k$--$\omega$ model is motivated by its near-wall formulation with a prescribed $\omega$ distribution, which makes analytical treatment more tractable than in the $k$--$\varepsilon$ model. In earlier work with a similar motivation \citep{joo2024reynolds,joo2025global}, the role of the buoyancy term in the $\varepsilon$ equation was examined within the $k$--$\varepsilon$ framework. However, in the standard $k$--$\varepsilon$ model, the boundary conditions of the $\varepsilon$ equation depend on spatial gradients of $k$ and $\varepsilon$, so a unique steady-state solution is not strictly guaranteed. In a closed Rayleigh--B\'enard configuration, the predicted $k$ and $\varepsilon$ can exhibit exponential growth over time depending on the initial conditions \citep{joo2024reynolds,joo2025global}.

This difficulty does not arise in the $k$--$\omega$ formulation. In the $k$--$\omega$ model, $\omega$ attains a peak at the wall and its near-wall distribution is prescribed solely by viscosity \citep{wilcox2006turbulence}. Consequently, even in a Rayleigh--B\'enard configuration, the steady-state $\omega$ distribution is uniquely determined, enabling an explicit analytical solution. Establishing such an analytical solution for the $k$--$\omega$ model is therefore one of the central objectives of the present study. Because of the distinct mathematical structures of the two models, the analytical approach introduced here for the $k$--$\omega$ model is largely independent of the prior analysis for the $k$--$\varepsilon$ model.

The main contributions of this study are threefold: 
(i) an analytical framework that clarifies how the scale-determining equation governs buoyancy-dominated turbulence and its simulation behavior; 
(ii) a closed-form analytical solution for statistically one-dimensional Rayleigh--B\'enard convection under the standard $k$--$\omega$ closure, yielding explicit $\Nu$--$\Ra$--$\Pran$ scalings; and 
(iii) corrections to buoyancy-related model terms designed to bring the RANS closure into closer agreement with established experimental trends across broad $\Ra$ and $\Pran$ ranges.

The remainder of the paper is organized as follows. 
\S\ref{sec:2} summarizes previous treatments of buoyancy in two-equation closures and presents the standard $k$--$\omega$ formulation of \citet{wilcox2006turbulence}. 
\S\ref{sec:3} derives an analytical solution for statistically one-dimensional Rayleigh--B\'enard convection, yielding an explicit relation for $\Nu$ as a function of $\Ra$ and $\Pran$. 
\S\ref{sec:4} proposes a new buoyancy-correction methodology based on the established analytical solution. The buoyancy term in the scale equation is modified, and a buoyancy-wall model is introduced into the turbulent thermal diffusivity. The resulting corrected $\omega$ equation is formulated by synthesizing the established analytical criterion for negative buoyant production with the new analytical criterion derived here for positive buoyant production. 
\S\ref{sec:5} further validates the proposed model beyond the one-dimensional Rayleigh--B\'enard test case through numerical simulations of two types of internally heated convection, unstably stratified Couette flow, Rayleigh--B\'enard convection in a two-dimensional enclosure, and vertically heated natural convection with varying aspect ratios. 
Finally, \S\ref{sec:6} provides a discussion, and \S\ref{sec:7} concludes the paper.

\section{Background and standard $k$--$\omega$ formulation}\label{sec:2}
\subsection{Previous treatments of buoyancy in two-equation RANS models}\label{sec:2.2}

This subsection reviews how buoyancy effects have been incorporated into the $\varepsilon$- and $\omega$-equations. Because most previous work has been formulated within the $k$--$\varepsilon$ framework, that formulation is outlined first, followed by its relation to the $k$--$\omega$ model.

At the outset, buoyancy in incompressible flow is represented using the Boussinesq approximation. The Reynolds-averaged momentum and temperature transport equations are then written as
\begin{equation}
	\frac{\partial U_i}{\partial t} 
	+ U_j \frac{\partial U_i}{\partial x_j} 
	= - \frac{\partial P}{\partial x_i} 
	- g_i b (T - T_0) 
	+ \frac{\partial}{\partial x_j} 
	\left( \nu \frac{\partial U_i}{\partial x_j} - \overline{u_i u_j} \right),
\end{equation}
\begin{equation}\label{eq: T eq}
	\frac{\partial T}{\partial t} 
	+ U_i \frac{\partial T}{\partial x_i} 
	= Q
	+ \frac{\partial}{\partial x_i} 
	\left( a \frac{\partial T}{\partial x_i} - \overline{\theta u_i} \right),
\end{equation}
where $U_i$ and $u_i$ denote the mean and fluctuating velocity components, respectively, and $T$ and $\theta$ are the mean and fluctuating temperature. 
$x_i$ are the Cartesian coordinates and $t$ denotes time. 
$P$ is the mean kinematic pressure, $g_i$ is the gravitational acceleration, $b$ is the thermal expansion coefficient, $T_0$ is a reference temperature, $\nu$ is the kinematic viscosity, $Q$ is the volumetric heat source per unit heat capacity, and $a$ is the thermal diffusivity.
When a lower wall is present, the choice of $T_0$ merely shifts the hydrostatic reference state and does not influence the fluid motion; for simplicity, $T_0$ is therefore set to zero.

The overbar denotes Reynolds averaging. The Reynolds stress tensor $\overline{u_i u_j}$ and the turbulent heat flux vector $\overline{\theta u_i}$ are modeled using the turbulent-viscosity and gradient-diffusion hypotheses as
\begin{equation}
	\overline{u_i u_j} = \frac{2}{3}k\delta_{ij} 
	- 2\nu_T S_{ij}, 
	\qquad
	\overline{\theta u_i} = - a_T \frac{\partial T}{\partial x_i},
\end{equation}
where
\begin{equation} \label{eq: aT}
	S_{ij} = \frac{1}{2}
	\left( 
	\frac{\partial U_i}{\partial x_j} + 
	\frac{\partial U_j}{\partial x_i} 
	\right),
	\qquad 
	a_T = \frac{\nu_T}{\Pran_T}.
\end{equation}
Here $\delta_{ij}$ is the Kronecker delta, $S_{ij}$ is the mean strain-rate tensor, $\nu_T$ is the turbulent viscosity, $a_T$ is the turbulent thermal diffusivity, and the turbulent Prandtl number is taken as $\Pran_T = 0.89$.

When the standard $k$--$\varepsilon$ model of \citet{launder1974application} is extended to include buoyant production, it takes the form
\begin{equation}
	\nu_T = C_{\mu} f_{\mu} \frac{k^{2}}{\varepsilon},
\end{equation}
\begin{equation}
	\frac{\partial k}{\partial t}
	+ U_j \frac{\partial k}{\partial x_j}
	=
	\mathcal{P}
	+ \mathcal{P}_b
	- \varepsilon
	- 2\nu 
	\left( 
	\frac{\partial k^{1/2}}{\partial x_j} 
	\frac{\partial k^{1/2}}{\partial x_j} 
	\right) 
	+ 
	\frac{\partial}{\partial x_j}
	\left[
	\left( \nu + \frac{\nu_T}{\sigma_k} \right) 
	\frac{\partial k}{\partial x_j}
	\right],
\end{equation}
\begin{equation}\label{eq:k-e, epsilon eq}
	\frac{\partial \varepsilon}{\partial t}
	+ U_j \frac{\partial \varepsilon}{\partial x_j}
	=
	C_{\varepsilon 1} \frac{\varepsilon}{k} 
	\left( \mathcal{P} + C_{\varepsilon b}\mathcal{P}_b \right) 
	- 
	C_{\varepsilon 2} f_{\varepsilon} 
	\frac{\varepsilon^{2}}{k}
	+ 
	\frac{\partial}{\partial x_j}
	\left[
	\left( \nu + \frac{\nu_T}{\sigma_\varepsilon} \right)
	\frac{\partial \varepsilon}{\partial x_j}
	\right],
\end{equation}
where the turbulent kinetic energy production terms due to the mean velocity gradient and buoyancy are
\begin{equation}
	\mathcal{P} = -\overline{u_i u_j} \frac{\partial U_i}{\partial x_j},
	\quad \mathrm{and} \quad
	\mathcal{P}_b = -  g_i b \overline{\theta u_i } .
\end{equation}
The damping functions are
\begin{equation}
	f_{\varepsilon} = 1.0 - 0.3\exp 
	\!\left[-\left(\frac{k^2}{\nu\varepsilon}\right)^{2}\right],
	\qquad
	f_{\mu} = 
	\exp\!\left[
	-\frac{3.4}{1+0.02k^2/(\nu\varepsilon)}
	\right].
\end{equation}
The closure coefficients are
$C_{\varepsilon 1} = 1.44$,
$C_{\varepsilon 2} = 1.92$,
$C_{\mu} = 0.09$,
$\sigma_k = 1.0$,
and $\sigma_\varepsilon = 1.3$.
The boundary conditions are $k = \varepsilon = 0$ at no-slip walls.

To establish correspondence with the $k$--$\omega$ formulation, the $\varepsilon$-equation can be reformulated by introducing $\varepsilon=\beta^{\ast}k\omega$ and $\nu_T \approx k/\omega$, as demonstrated in \citet{pope2000turbulent}, where $\beta^{\ast}$ is a model constant. 
Assuming $\nu_T \gg \nu$, such that molecular viscosity is negligible and $f_{\mu}=f_{\varepsilon}=1$, substitution yields
\begin{eqnarray}
	\frac{\partial \omega}{\partial t}
	+ U_j \frac{\partial \omega}{\partial x_j}
	&=&
	\left( C_{\varepsilon 1} - 1\right)  
	\frac{\omega}{k}\mathcal{P}
	+ 
	\left( C_{\varepsilon 1}C_{\varepsilon b} - 1 \right)
	\frac{\omega}{k}\mathcal{P}_b
	- 
	\left( C_{\varepsilon 2} -1 \right)\omega^{2}
	\nonumber \\[3pt]
	&& \mbox{}
	+ 
	\frac{C_{\mu}}{\beta^{\ast}}
	\left( \frac{1}{\sigma_\varepsilon} - \frac{1}{\sigma_k} \right) 
	\!\left( 
	\frac{1}{k}\frac{\partial k}{\partial x_j}
	\frac{\partial k}{\partial x_j}
	+ 
	\frac{\partial^2 k}{\partial x_j \partial x_j} 
	\right) \\[3pt]
	&& \mbox{}
	+ 
	\frac{C_{\mu}}{\beta^{\ast}}
	\left( \frac{1}{\sigma_\varepsilon} + \frac{1}{\sigma_k} \right) 
	\frac{1}{\omega}
	\frac{\partial k}{\partial x_j}
	\frac{\partial \omega}{\partial x_j}
	+  
	\frac{\partial }{\partial x_j}
	\!\left[
	\frac{\nu_T}{\sigma_{\varepsilon}} 
	\frac{\partial \omega}{\partial x_j}
	\right].
	\nonumber
\end{eqnarray}
The standard $k$--$\omega$ formulation contains corresponding model terms,
\[
\frac{\partial \omega}{\partial t}
+ U_j \frac{\partial \omega}{\partial x_j} = \alpha \frac{\omega}{k} (\mathcal{P} + C_{\omega b} \mathcal{P}_b) + \cdots,
\]
where $\alpha=0.52$ and $C_{\omega b}$ are model constants.  
In conjunction with this, the following approximate correspondence is commonly adopted:
\begin{equation}\label{eq: Cwb Ceb}
	\frac{C_{\varepsilon 1}C_{\varepsilon b} - 1 }{C_{\varepsilon 1} - 1} 
	\approx
	\frac{(\alpha + 1)C_{\varepsilon b} - 1 }{\alpha}
	\approx
	C_{\omega b}.
\end{equation}
The full formulation of the $k$--$\omega$ model is presented in \S\ref{sec:2.1}.

In most prior studies, discussions of buoyant production have been anchored to the $k$--$\varepsilon$ framework. 
As noted in \S\ref{sec:1}, there is no universally accepted standard for $C_{\varepsilon b}$; reported values range approximately from 0 to 1, as documented in \citet{hanjalic2011modelling,burchard2007applied,henkes1991natural,peng1999computation,hanjalic1993computation}. 
The proposal of \citet{markatos1984laminar} to omit the buoyancy contribution corresponds to $C_{\varepsilon b}=0$. 
An equally common practice is to treat shear and buoyant production on the same footing, corresponding to $C_{\varepsilon b}=1$, as in \citet{ince1989computation,craft1996recent,choi2012turbulence}. 

Within this spectrum, the reference work of \citet{rodi1980turbulence} suggested that $C_{\varepsilon b}$ should approach unity in vertical buoyant shear layers and vanish in horizontal layers. 
Along similar lines, \citet{henkes1991natural} proposed a velocity-ratio form for the buoyancy coefficient, $C_{\varepsilon b}=\tanh|v/u|$, where $v$ denotes the velocity component aligned with gravity and $u$ the component perpendicular to it. 
Another approach by \citet{rodi1987examples} recommended $C_{\varepsilon b}=1$ for unstable stratification and near zero for stable layers, a prescription based on an optimization for an unstable shear flow at a reduced Froude number of 0.9 (defined as the ratio of the maximum shear-velocity difference to the buoyancy free-fall velocity), as reported by \citet{viollet1987modelling}.

These empirical prescriptions have propagated into widely used turbulence solvers, both open-source and commercial, but they are employed in mutually inconsistent ways. 
OpenFOAM \citep{openfoam2025} adopts $C_{\varepsilon b}=1$ by default, whereas Fluent \citep{fluent2013ansys} uses $C_{\varepsilon b}=0$ and offers the alternative $C_{\varepsilon b}=\tanh|v/u|$, following \citet{henkes1991natural}. 
In CFX \citep{cfx2006theory}, the buoyancy source term in the $\omega$-equation is implemented as
\[
\frac{\omega}{k} 
\left[
(\alpha + 1) \max(\mathcal{P}_b,0)  f_{\phi} - \mathcal{P}_b
\right],
\]
which corresponds to $C_{\omega b}=1$ for $\mathcal{P}_b>0$ and $C_{\omega b}\approx-2$ for $\mathcal{P}_b<0$, with the default setting $f_{\phi}=1$. 
This formulation is identical to the recommendation of \citet{rodi1987examples} for $C_{\varepsilon b}$, as related through Eq.~(\ref{eq: Cwb Ceb}). 
An alternative is provided as $f_{\phi}=\sin(\phi)$, where $\phi$ denotes the angle between the flow direction and gravity, again following \citet{henkes1991natural}.

However, both Henkes’ proposal and the introduction of $\sin(\phi)$ in CFX violate Galilean invariance. 
A simple counterexample arises by shifting the reference frame with velocity $v^{\prime}$ in the direction of gravity; in this case $C_{\varepsilon b}=\tanh|(v-v^{\prime})/u|$ changes its value. 
Even if this were interpreted as an absolute-frame velocity relative to a stationary wall, it would imply that a locally defined model equation depends on a remote boundary without explicit specification. 
From any perspective, such approaches undermine the formal completeness of two-equation closures.

Across the cited studies \citep{rodi1980turbulence,markatos1984laminar,rodi1987examples,viollet1987modelling,henkes1991natural,hanjalic1993computation,peng1999computation,burchard2007applied,hanjalic2011modelling, hanjalic2021reassessment}, 
it has been repeatedly emphasized that, despite various empirical proposals, no firm analytical criterion exists for incorporating buoyancy effects into the scale-determining equations. 
Moreover, even when buoyancy effects are introduced in an analogous manner in the $k$--$\varepsilon$ and $k$--$\omega$ models, there is no guarantee that the resulting behavior will remain consistent between the two formulations. 
These limitations motivate the analytical framework developed in the remainder of this paper.

\subsection{Wilcox (2006) $k$--$\omega$ model and buoyancy extensions}\label{sec:2.1}

This study analyzes the Wilcox (2006) $k$--$\omega$ model \citep{wilcox2006turbulence}. 
The notation for the model variables and constants follows that reference exactly; only the buoyancy-related terms are added. 
The full set of model equations, including the transport equations for $k$ and $\omega$, is given by
\begin{equation}
	\nu_T = \frac{k}{\tilde{\omega}}, \qquad 
	\tilde{\omega} = \max \left\lbrace \omega, 	
	C_{lim} \sqrt{\frac{2S_{ij}S_{ij}}{\beta^{\ast}}} \right\rbrace ,
\end{equation}
\begin{equation}
	\frac{\partial k}{\partial t}
	+ U_j \frac{\partial k}{\partial x_j}
	=
	\mathcal{P} + \mathcal{P}_b
	-\beta^{\ast} k \omega
	+ \frac{\partial }{\partial x_j}
	\left[\left( \nu + \sigma^{\ast} \frac{k}{\omega} \right) 
	\frac{\partial k}{\partial x_j}  \right] ,
\end{equation}
\begin{equation}\label{eq: omega eq}
	\frac{\partial \omega}{\partial t}
	+ U_j \frac{\partial \omega}{\partial x_j}
	=
	\alpha \frac{\omega}{k} \left( \mathcal{P} + C_{\omega b} \mathcal{P}_b \right) 
	-\beta_0 f_{\beta} \omega^{2}
	+ \frac{\sigma_d}{\omega} 
	\frac{\partial k}{\partial x_j} 
	\frac{\partial \omega}{\partial x_j}
	+ \frac{\partial }{\partial x_j}
	\left[\left( \nu + \sigma \frac{k}{\omega} \right) 
	\frac{\partial \omega}{\partial x_j}  \right] .
\end{equation}

The auxiliary relations are defined as
\begin{equation}
	\sigma_d = 
	\left\{
	\begin{array}{ll}
		0, & \frac{\partial k}{\partial x_j} 
		\frac{\partial \omega}{\partial x_j} \le 0 \\[6pt]
		\sigma_{do}, & \frac{\partial k}{\partial x_j} 
		\frac{\partial \omega}{\partial x_j} > 0 \:,
	\end{array}
	\right. 
\end{equation}
\begin{equation}
	f_{\beta} = \frac{1+ 85 \chi_\omega}{1+ 100 \chi_\omega}, \qquad
	\chi_\omega = 
	\left| 
	\frac{\Omega_{ij}\Omega_{jk}S_{ki}}{(\beta^{\ast}\omega)^{3}} 
	\right| ,
\end{equation}
with
\begin{equation}
	\Omega_{ij} = \frac{1}{2}
	\left( 
	\frac{\partial U_i}{\partial x_j} 
	- 
	\frac{\partial U_j}{\partial x_i} 
	\right).
\end{equation}

The closure coefficients are 
$C_{lim}=0.875$, 
$\beta^{\ast}=0.09$, 
$\sigma^{\ast}=0.6$,
$\alpha=0.52$,
$\beta_0=0.0708$,
$\sigma=0.5$, and
$\sigma_{do}=0.125$.  
In the $\omega$-equation, the model constant associated with buoyancy effects is $C_{\omega b}$.

At no-slip walls, the boundary conditions are
\begin{equation}\label{eq: no slip BC}
	k=0, \qquad
	\omega = \frac{6\nu}{\beta_0 n^2},
\end{equation}
where, for a cell-centered discretization of $\omega$, $n$ denotes the wall-normal distance from the wall to the center of the first off-wall grid cell.

\section{Analytical solution in Rayleigh--B\'enard convection}\label{sec:3}

The objective of this analysis is to establish how the standard $k$--$\omega$ model predicts the Nusselt number as a function of the Rayleigh and Prandtl numbers in a buoyancy-driven flow. 
As noted in \S\ref{sec:1}, a statistically one-dimensional Rayleigh--B\'enard configuration is selected as the canonical test case, in which all mean quantities depend solely on the vertical coordinate $z$ and the mean velocity vanishes. 
In this configuration, the turbulent kinetic energy budget is governed entirely by buoyant production, thereby isolating the role of the buoyancy term in the $\omega$ equation.

\begin{figure}[h]
	\centering
	\begin{subfigure}{0.3\textwidth}
		\centering
		\includegraphics[scale=1.0]{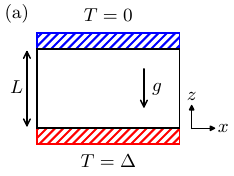}
	\end{subfigure}
	\begin{subfigure}{0.3\textwidth}
		\centering
		\includegraphics[scale=1.0]{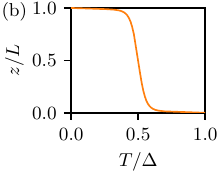}
	\end{subfigure}
	\begin{subfigure}{0.3\textwidth}
		\centering
		\includegraphics[scale=1.0]{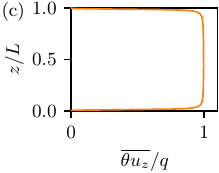}
	\end{subfigure}
	\caption{(a) Rayleigh--B\'enard convection setup, (b) mean temperature distribution, and (c) turbulent heat flux.}
	\label{fig:1}
\end{figure}

Figure~\ref{fig:1}(a) depicts the Rayleigh--B\'enard configuration and its boundary conditions. 
The walls are no-slip and isothermal, with $T=0$ at $z=L$ and $T=\Delta$ at $z=0$. 
The dimensionless control parameters are the Rayleigh and Prandtl numbers, defined as 
$\Ra = g b \Delta L^{3}/(\nu a)$ and $\Pran = \nu/a$. 
The Nusselt number is defined as
\begin{equation}\label{eq: Nu def}
	\Nu = \frac{qL}{a \Delta},
	\qquad
	q = -a \frac{\mathrm{d} T}{\mathrm{d} z} + \overline{\theta u_z},
\end{equation}
where the total heat flux $q$ is the sum of the conductive and turbulent contributions and remains constant in the vertical direction. 
Figures~\ref{fig:1}(b,c) illustrate the vertical variation of the mean temperature and turbulent heat flux. 
Gravity acts in the negative $z$ direction, with $U_i=0$, $\mathrm{d}T/\mathrm{d}z<0$, and $\overline{\theta u_z}>0$.

In this study, the kinetic and thermal boundary-layer thicknesses, $\lambda_\nu$ and $\lambda_a$, are defined as the locations where $\nu=\nu_T$ and $a=a_T$, respectively. 
Because the diffusion terms in the $k$- and $\omega$-equations depend on both $\nu$ and $\nu_T$, the contrast between their wall and bulk distributions must be interpreted separately with respect to $\lambda_\nu$. 
By contrast, $\lambda_a$ characterizes the heat-transfer structure: within $\lambda_a$, the conductive approximation $q \approx -a\, \mathrm{d}T/\mathrm{d}z$ holds adjacent to the wall, whereas outside $\lambda_a$ the turbulent relation $q \approx \overline{\theta u_z}$ applies.

\subsection{Distribution of $\omega$ and $k$}\label{sec:3.1}

The distribution of $\omega$ is established first. 
Owing to the symmetry of the domain, only the half region $0 \le z \le L/2$ is considered. 
For the analytical derivation, the governing equation is simplified uniformly under the condition $\nu \gg \nu_T$ in the region $z \le \lambda_\nu$, whereas in the region $z \ge \lambda_\nu$ it is simplified under the opposite condition $\nu \ll \nu_T$. 
Under these assumptions, the resulting $\omega$ distributions are continuous at the interface $z = \lambda_\nu$, although their spatial gradients are discontinuous.

Close to the wall ($z \le \lambda_\nu$), the distribution of $\omega$ is fixed by the boundary condition of the model \citep{wilcox2006turbulence}:
\begin{equation}\label{eq: omega BL}
	\omega = \frac{6\nu}{\beta_0 z^2}.
\end{equation}
This expression can also be recovered directly from the $\omega$-equation by balancing the sink term with viscous diffusion.

In the bulk region, $\lambda_\nu \le z \le L/2$, the distribution of $\omega$ is determined under the approximation $\nu_T \gg \nu$. 
Using the heat balance of Eq.~(\ref{eq: Nu def}), such that $-(a+a_T)(\mathrm{d}T/\mathrm{d}z)=q$, the buoyant production term becomes
\begin{equation}\label{eq: Pb bulk}
	\mathcal{P}_b = g b \overline{\theta u_z}
	= - g b a_T \frac{\mathrm{d} T}{\mathrm{d} z}
	= \frac{g b q}{1+a/a_T} \approx g b q .
\end{equation}
Here, the assumption $a_T \gg a$, implying $q \approx \overline{\theta u_z}$, is invoked to admit a simple closed-form solution; retaining the full expression would render the equations substantially more complicated. 
Because the objective of this analysis is to obtain structural insight into the model through a tractable analytical formulation, the study is restricted to this approximation. 
For sufficiently large Nusselt numbers, this simplification is well justified, as illustrated in Fig.~\ref{fig:1}(c).

In the bulk region, it is assumed that production in the $k$-equation is locally balanced by dissipation. 
Combining $\varepsilon = \beta^{\ast} k \omega$, $\mathcal{P}_b = g b q$, and $\mathcal{P}_b = \varepsilon$ yields
\begin{equation}\label{eq: k to omega at bulk}
	k = \frac{g b q}{\beta^{\ast}\omega}.
\end{equation}

Since production and dissipation balance in the $k$-equation but not in the $\omega$-equation, diffusion must be retained in the latter. 
The $\omega$-equation in the bulk then reduces to
\begin{equation}
	0 = \alpha \frac{\omega}{k} C_{\omega b} (gbq)
	- \beta_0 \omega^2
	+ \frac{\mathrm{d}}{\mathrm{d}z}
	\left(\sigma \frac{k}{\omega} \frac{\mathrm{d}\omega}{\mathrm{d}z}\right).
\end{equation}

Substituting $k$ from Eq.~(\ref{eq: k to omega at bulk}) leads to
\begin{equation} \label{eq: omega bulk diff}
	\frac{\mathrm{d}}{\mathrm{d}z}
	\left(\frac{1}{\omega^2}\frac{\mathrm{d}\omega}{\mathrm{d}z}\right)
	=
	\frac{\beta^{\ast}}{\sigma g b q}
	\left(\beta_0-\alpha\beta^{\ast}C_{\omega b}\right)\omega^2 .
\end{equation}
Because $\omega$ is largest at the wall and decreases toward the domain center, the right-hand side must be positive, which imposes the constraint
\begin{equation}
	C_{\omega b} < \frac{\beta_0}{\alpha \beta^{\ast}} = 1.52 
	\qquad \text{for} \qquad \mathcal{P}_b>0 .
\end{equation}

Introducing the dimensionless variables
\begin{equation} \label{eq:tau definition}
	\tau = \frac{\nu}{L^2 \omega}, \qquad
	z^{\ast} = \frac{z}{L},
\end{equation}
the $\omega$-equation can be written as
\begin{equation} \label{eq:tau}
	\frac{\mathrm{d}^2 \tau}{\mathrm{d}{z^{\ast}}^2}
	= -\frac{A}{\tau^2},
\end{equation}
where
\begin{equation}
	A =
	\frac{\beta^{\ast}}{\sigma}
	\left(\beta_0-\alpha\beta^{\ast}C_{\omega b}\right)
	\Pran^{2}\Nu^{-1}\Ra^{-1}.
\end{equation}

The corresponding boundary conditions are
\begin{equation}
	\tau=\tau_\nu=
	\frac{\beta_0}{6}\left(\frac{\lambda_\nu}{L}\right)^2
	\quad\text{at}\quad
	z^{\ast}=\frac{\lambda_\nu}{L},
\end{equation}
\begin{equation}
	\frac{\mathrm{d}\tau}{\mathrm{d}z^{\ast}}=0
	\quad\text{at}\quad
	z^{\ast}=\frac{1}{2},
\end{equation}
where $\tau_\nu$ denotes the value of $\tau$ at $z^{\ast}=\lambda_\nu/L$.

Multiplying Eq.~(\ref{eq:tau}) by $\mathrm{d}\tau/\mathrm{d}z^{\ast}$ and integrating under these boundary conditions gives
\begin{equation} \label{eq:tau2}
	\left(\frac{\mathrm{d}\tau}{\mathrm{d}z^{\ast}}\right)^2
	=2A\left(\frac{1}{\tau}-\frac{1}{\tau_c}\right),
\end{equation}
where $\tau_c$ denotes the value of $\tau$ at the center of the domain ($z^{\ast}=1/2$).

A second integration yields
\begin{equation}\label{eq: tau-z}
	z^{\ast} =
	\frac{1}{2} -
	\sqrt{\frac{\tau_c^{3}}{2A}}
	\left[
	\arccos\!\left(\sqrt{\frac{\tau}{\tau_c}}\right)
	+ \sqrt{\frac{\tau}{\tau_c}\!\left(1-\frac{\tau}{\tau_c}\right)}
	\right],
\end{equation}
which satisfies $z^{\ast}=1/2$ at $\tau=\tau_c$.

The value of $\tau_c$ follows from imposing $\tau=\tau_\nu$ at $z^{\ast}=\lambda_\nu/L$. 
When the boundary layer is thin, $\omega$ is very large near the wall and much smaller in the core, so that $\tau_\nu/\tau_c \ll 1$. 
Expanding the $\arccos$ and square-root terms in a Taylor series gives
\begin{eqnarray}\label{eq: lambda_nu}
	\frac{\lambda_\nu}{L} =
	\frac{1}{2} -
	\sqrt{\frac{\tau_c^{3}}{2A}}
	\left[
	\frac{\pi}{2}
	-\frac{2}{3}\!\left(\frac{\tau_\nu}{\tau_c}\right)^{3/2}
	+\cdots
	\right],
\end{eqnarray}
and hence
\begin{eqnarray}\label{eq: tau1}
	\tau_c =
	\left(\frac{2A}{\pi^{2}}\right)^{1/3}
	\left[
	1
	-\frac{4}{3}\frac{\lambda_\nu}{L}
	+\frac{4}{9}\sqrt{\frac{2}{A}}\tau_\nu^{3/2}
	+\cdots
	\right].
\end{eqnarray}
Thus, $\omega$ attains its minimum value at $z=L/2$, given by $\nu L^{-2}\tau_c^{-1}$.

In the bulk region ($\lambda_\nu \le z \le L/2$), the distribution of $k$ follows directly from Eq.~(\ref{eq: k to omega at bulk}) by substituting the previously obtained $\omega$ profile (Eq.~\ref{eq: tau-z}). 
In the near-wall region ($z \le \lambda_\nu$), the $k$ distribution is determined by the intrinsic behavior of the $k$--$\omega$ model \citep{wilcox2006turbulence}. 
The $k$-equation reduces to a balance between dissipation and viscous diffusion,
\begin{equation}
	\beta^{\ast}k\omega = \nu \frac{\mathrm{d}^2k}{\mathrm{d}z^2},
\end{equation}
and substituting $\omega = 6\nu/(\beta_0 z^2)$ gives
\begin{equation}
	\frac{6\beta^{\ast}}{\beta_0}\frac{k}{z^2}
	= \frac{\mathrm{d}^2k}{\mathrm{d}z^2}.
\end{equation}
Assuming a power-law form $k\propto z^p$ yields
\begin{equation}\label{eq: p of k}
	p = \frac{1}{2} +
	\sqrt{\frac{6\beta^{\ast}}{\beta_0}+\frac{1}{4}}
	\approx 3.307 .
\end{equation}
Hence, for $z\le\lambda_\nu$, the near-wall distribution of $k$ is approximated as
\begin{equation}
	k = k_\nu\left(\frac{z}{\lambda_\nu}\right)^p,
\end{equation}
where $k_\nu$ denotes the value of $k$ at $z=\lambda_\nu$.

Finally, $k_\nu$ is obtained by enforcing continuity between the near-wall and bulk solutions at $z=\lambda_\nu$, which gives
\begin{equation}\label{eq: k0}
	k_\nu
	= \frac{g b q}{\beta^{\ast}}
	\frac{\beta_0 \lambda_\nu^2}{6\nu}.
\end{equation}

\subsection{Boundary-layer thicknesses}\label{sec:3.3}

The kinetic boundary-layer thickness $\lambda_\nu$ is defined as the location where $\nu_T=\nu$. 
Using Eq.~(\ref{eq: omega BL}), the value of $\omega$ at $z=\lambda_\nu$ is 
$\omega = 6\nu/(\beta_0 \lambda_{\nu}^{2})$. 
Substituting this into Eq.~(\ref{eq: k0}), the turbulent viscosity $\nu_T = k/\omega$ evaluated at 
$z \le \lambda_\nu$ becomes
\begin{equation}\label{eq: nuT BL}
	\nu_T
	= \frac{g b q}{\beta^{\ast}}
	\left( \frac{6\nu}{\beta_0 \lambda_\nu^{2}} \right)^{-2} 
	\left( \frac{z}{\lambda_\nu}\right)^{p+2}.
\end{equation}
Imposing the defining condition $\nu = \nu_T$ at $z=\lambda_\nu$ yields the estimate
\begin{equation}\label{eq: kinetic BL}
	\frac{\lambda_\nu}{L} = 
	\left( \frac{36 \beta^{\ast}}{\beta_0^{2}} \right)^{1/4} 
	\Nu^{-1/4} \Ra^{-1/4} \Pran^{1/2},
\end{equation}
and consequently
\begin{equation}
	\tau_\nu = {\beta^{\ast}}^{1/2} 
	\Nu^{-1/2} \Ra^{-1/2} \Pran,
\end{equation}
which represents the value of $\omega = \nu /(L^2 \tau_\nu)$ at $z=\lambda_\nu$.

When $\Pran \ge \Pran_T$, and hence $\lambda_a \le \lambda_\nu$, the thermal boundary-layer thickness is estimated from the condition $a=a_T$ at $z=\lambda_a$, using the near-wall distribution of $\nu_T/\nu$:
\begin{equation}\label{eq: aT a}
	\frac{a_T}{a} = 
	\frac{\Pran}{\Pran_T}\frac{\nu_T}{\nu} = 
	\frac{\Pran}{\Pran_T} \left( \frac{\lambda_a}{\lambda_\nu} \right)^{p+2} = 1,
\end{equation}
where $p = 3.307$. 
It then follows that
\begin{equation}\label{eq: thermal BL high Pr}
	\frac{\lambda_a}{L} = 
	\left( \frac{36 \beta^{\ast}}{\beta_0^{2}} \right)^{1/4}  
	\Pran_T^{1/(p+2)}
	\Nu^{-1/4} \Ra^{-1/4} \Pran^{p/(2p+4)}.
\end{equation}

On the other hand, when $\Pran \le \Pran_T$, so that $\lambda_a \ge \lambda_\nu$, $\lambda_a$ must be determined from the bulk distribution of $\omega$. 
The condition $a=a_T$, with $\nu_T = k/\omega \approx g b q {\beta^{\ast}}^{-1} (L^2 \tau_a / \nu)^2$, leads to
\begin{equation}\label{eq: tau_a bulk}
	\tau_a = {\beta^{\ast}}^{1/2} \Pran_T^{1/2} \Nu^{-1/2} \Ra^{-1/2} \Pran^{1/2}, 
\end{equation}
where $\tau_a$ denotes the value of $\tau$ at $z=\lambda_a$. 
Assuming $\tau_a \ll \tau_c$, a Taylor expansion of the $\tau$--$z^{\ast}$ relation (Eqs.~\ref{eq: tau-z} and \ref{eq: lambda_nu}) gives
\begin{eqnarray}
	\frac{\lambda_a}{L} \approx \frac{1}{2} - \sqrt{\frac{\tau_c^3}{2A}}
	\left[ \frac{\pi}{2} - \frac{2}{3} \left( \frac{\tau_a}{\tau_c} \right)^{3/2} + \cdots \right],
\end{eqnarray}
and $\lambda_a$ for $\Pran \le \Pran_T$ is obtained as
\begin{eqnarray}\label{eq: thermal BL low Pr}
	\frac{\lambda_a - \lambda_\nu}{L}
	&\approx& \frac{2}{3} \sqrt{\frac{\tau_c^3}{2A}}
	\left[ \left( \frac{\tau_a}{\tau_c} \right)^{3/2} - \left( \frac{\tau_\nu}{\tau_c} \right)^{3/2} \right] \nonumber\\
	&=&
	\frac{\sqrt{2}}{3} \frac{\sigma^{1/2}}{\alpha^{1/2} {\beta^{\ast}}^{1/4}} 
	\left( \frac{\beta_0}{\alpha \beta^{\ast}} - C_{\omega b} \right)^{-1/2}
	\left( \Pran_T^{3/4} - \Pran^{3/4} \right) 
	\Nu^{-1/4} \Ra^{-1/4} \Pran^{-1/4}.
	\nonumber\\
\end{eqnarray}

By substituting the expressions for the boundary-layer thicknesses derived above, the distributions of $\omega$ and $k$ can be expressed explicitly as functions of $\Ra$, $\Pran$, and $\Nu$.

\subsection{Temperature profile}\label{sec:3.4}

The temperature profile is analyzed separately in the near-wall region ($z \le \lambda_\nu$) and in the bulk region ($\lambda_\nu \le z \le L/2$), providing the basis for establishing the relation among $\Nu$, $\Ra$, and $\Pran$ discussed in \S\ref{sec:3.5}. 
The associated heat balance is governed by the thermal boundary-layer thickness $\lambda_a$. 
For $z > \lambda_a$, the heat flux can be approximated by the turbulent relation $q \approx \overline{\theta u_z}$, whereas for $z < \lambda_a$ the effect of molecular thermal diffusivity $a$ must also be retained.

\subsubsection{Near-wall contribution}\label{sec:3.4.1}

The near-wall temperature rise across the kinetic boundary layer, $\Delta_{BL}$, is first evaluated as
\begin{eqnarray}\label{eq: Delta_BL original}
	\frac{\Delta_{BL}}{\Delta} 
	&=& - \frac{1}{\Delta} \int_{0}^{\lambda_\nu}
	\frac{\mathrm{d} T}{\mathrm{d} z}  \mathrm{d} z 
	= \Nu \left( \frac{\lambda_\nu}{L} \right) \int_{0}^{1} 
	\frac{1}{1 + a_T / a}  \mathrm{d} z^{\prime} \nonumber \\[3pt]
	&=& \left( \frac{36 \beta^{\ast}}{\beta_0^{2}} \right)^{1/4} 
	\Nu^{3/4} \Ra^{-1/4} \Pran^{1/2}
	\int_{0}^{1}
	\frac{1}{1 + (\Pran / \Pran_T) {z^{\prime}}^{p^{\prime}}} \mathrm{d} z^{\prime},
\end{eqnarray}
where $p^{\prime} = p + 2 = 5.307$, $a_T / a = (\Pran/\Pran_T) (z/\lambda_\nu)^{p+2}$ (Eqs.~\ref{eq: aT} and \ref{eq: nuT BL}), and $z^{\prime} = z / \lambda_\nu$. 
The integral in the second line approaches unity when $\Pran / \Pran_T \ll 1$.

For $\Pran / \Pran_T \gg 1$, the asymptotic value of the integral can be obtained by applying the variable transformation $\eta = (\Pran / \Pran_T){z^{\prime}}^{p^{\prime}}$, yielding \citep{arfken2013mathematical}:
\begin{eqnarray}\label{eq: near wall temperature integral}
	\int_{0}^{1}
	\frac{1}{1 + (\Pran / \Pran_T){z^{\prime}}^{p^{\prime}}}   \mathrm{d} z^{\prime}
	&\approx&
	\frac{1}{p^{\prime}}
	\left( \frac{\Pran}{\Pran_T} \right)^{-1/p^{\prime}}
	\int_{0}^{\infty}
	\frac{\eta^{\frac{1}{p^{\prime}} - 1}}{1 + \eta}   \mathrm{d} \eta  \nonumber \\[3pt]
	&=&
	\frac{\pi}{p^{\prime} \sin (\pi / p^{\prime})}
	\left( \frac{\Pran}{\Pran_T} \right)^{-1/p^{\prime}} .
\end{eqnarray}

On the basis of this asymptotic behavior, the integral can be approximated by the algebraic expression
\[
\left[ 1 + 0.7535   (\Pran / \Pran_T)^{0.9338} \right]^{-1 / (0.9338 p^{\prime})},
\]
which satisfies the correct limits for both $\Pran / \Pran_T \ll 1$ and $\Pran / \Pran_T \gg 1$.  
This approximation yields a maximum relative error of 0.2\% over the range $10^{-4} \le (\Pran / \Pran_T) \le 10^{4}$.

Combining the above results, the near-wall temperature rise becomes
\begin{equation}\label{eq: Delta_BL}
	\frac{\Delta_{BL}}{\Delta} 
	=
	4.506  
	\Nu^{3/4} \Ra^{-1/4} \Pran^{1/2}
	\left[ 1 + 0.7535 \left( \frac{\Pran}{\Pran_T} \right)^{0.9338} \right]^{-1 / (0.9338 p^{\prime})}.
\end{equation}

\subsubsection{Bulk contribution for $\Pran \ge \Pran_T$}

The bulk contribution, $\Delta_{bulk}$, is now considered.  
For $\Pran \ge \Pran_T$, the approximation $\overline{\theta u_z} \approx q$ holds throughout the bulk region, where $z \ge \lambda_\nu \ge \lambda_a$.  
From the heat balance equation, the temperature gradient can be written as
\begin{equation}
	\frac{\mathrm{d} T}{\mathrm{d} z} 
	\approx - \frac{q}{\nu_T/\Pran_T}
	= - \frac{\Delta}{L} 
	\beta^{\ast} \Pran_T \Ra^{-1} \Pran \frac{1}{\tau^{2}} .
\end{equation}

Integrating between $z=\lambda_\nu$ and $z=L/2$ gives the bulk temperature rise:
\begin{eqnarray}\label{eq: Delta_bulk large Pr}
	\frac{\Delta_{bulk}}{\Delta} &\approx& 
	\frac{1}{\Delta}\int_{\lambda_\nu}^{L/2} \frac{q}{\nu_T / \Pran_T} \mathrm{d} z 
	= \beta^{\ast} \Pran_T \Ra^{-1} \Pran 
	\int_{\lambda_\nu/L}^{1/2} \frac{1}{\tau^{2}} \mathrm{d} z^{\ast} \nonumber \\[3pt]
	&=& \beta^{\ast} \Pran_T \Ra^{-1} \Pran 
	\int_{\tau_\nu}^{\tau_c} \frac{1}{\tau^{2}}\left[ 2A \left( \frac{1}{\tau} - \frac{1}{\tau_c} \right) \right] ^{-1/2} \mathrm{d} \tau \nonumber \\[3pt]
	&=& 
	\frac{\sqrt{2} \sigma^{1/2} {\beta^{\ast}}^{1/4} }{\left( \beta_0 - \alpha \beta^{\ast} C_{\omega b}   \right)^{1/2}}
	\left(1-\frac{\tau_\nu}{\tau_c}\right)^{1/2}
	\Nu^{3/4} \Ra^{-1/4} \Pran^{-1/2} \nonumber \\[3pt]
	&\approx& 
	2.532  
	\left( \frac{\beta_0}{\alpha \beta^{\ast}} - C_{\omega b} \right)^{-1/2}
	\Pran_T  
	\Nu^{3/4} \Ra^{-1/4} 
	\Pran^{-1/2},
\end{eqnarray}
where the final expression follows from the assumption $\tau_\nu/\tau_c \ll 1$.

\subsubsection{Bulk contribution for $\Pran \le \Pran_T$}\label{sec:3.4.3}

For $\Pran \le \Pran_T$, the ordering $\lambda_\nu \le \lambda_a$ holds.  
In the region $\lambda_\nu \le z \le \lambda_a$, the total thermal diffusivity $a + \nu_T/\Pran_T$ is less than $2a$, so that the approximation $\nu_T/\Pran_T \approx 0$ is adopted, whereas for $z > \lambda_a$ it is assumed that $\nu_T/\Pran_T \gg a$.  
This approximation modifies only the prefactor of the power law while leaving the exponent unchanged, which is sufficient for deriving analytical scaling laws.

The bulk temperature rise is written as
\begin{equation}
	\frac{\Delta_{bulk}}{\Delta} \approx
	\frac{1}{\Delta}\int_{\lambda_\nu}^{\lambda_a} \frac{q}{a} \mathrm{d} z 
	+
	\frac{1}{\Delta}\int_{\lambda_a}^{L/2} \frac{q}{\nu_T / \Pran_T} \mathrm{d} z .
\end{equation}

Using Eqs.~(\ref{eq: tau_a bulk}) and (\ref{eq: thermal BL low Pr}) gives
\begin{eqnarray}
	\frac{1}{\Delta}\int_{\lambda_\nu}^{\lambda_a} \frac{q}{a} \mathrm{d} z 
	&=& \Nu \left( \frac{\lambda_a - \lambda_\nu}{L} \right)  \nonumber \\[3pt]
	&\approx& 
	\frac{\sqrt{2}}{3} \frac{\sigma^{1/2}}{\alpha^{1/2}{\beta^{\ast}}^{1/4}}
	\left( \frac{\beta_0}{\alpha \beta^{\ast}} - C_{\omega b} \right)^{-1/2}
	\left( \Pran_T^{3/4} - \Pran^{3/4} \right) 
	\Nu^{3/4} \Ra^{-1/4}  \Pran^{-1/4},
\end{eqnarray}
and
\begin{eqnarray}
	\frac{1}{\Delta}\int_{\lambda_a}^{L/2} \frac{q}{\nu_T / \Pran_T} \mathrm{d} z 
	&=& \beta^{\ast} \Pran_T \Ra^{-1} \Pran 
	\left( \frac{2}{A \tau_a} \right)^{1/2} \left( 1 - \frac{\tau_a}{\tau_c} \right)^{1/2} \nonumber \\[3pt]
	&\approx&
	\sqrt{2}\frac{ \sigma^{1/2}}{{\alpha^{1/2} \beta^{\ast}}^{1/4} }
	\left( \frac{\beta_0}{\alpha \beta^{\ast}} - C_{\omega b}\right) ^{-1/2} 
	\Pran_T^{3/4} 
	\Nu^{3/4}
	\Ra^{-1/4}
	\Pran^{-1/4}.
\end{eqnarray}

Summing both contributions yields the total bulk temperature rise:
\begin{equation}\label{eq: Delta_bulk small Pr}
	\frac{\Delta_{bulk}}{\Delta} \approx
	2.532  
	\left( \frac{\beta_0}{\alpha \beta^{\ast}} - C_{\omega b}\right) ^{-1/2} 
	\left( \frac{4}{3}\Pran_T^{3/4} - \frac{1}{3}\Pran^{3/4} \right) 
	\Nu^{3/4} \Ra^{-1/4} \Pran^{-1/4}.
\end{equation}

\subsection{Dependence of $\Nu$ on $\Ra$ and $\Pran$}\label{sec:3.5}

The total temperature rise across the half-domain satisfies $\Delta/2=\Delta_{BL}+\Delta_{bulk}$.  
By combining the near-wall and bulk contributions obtained in Eqs.~(\ref{eq: Delta_BL}), (\ref{eq: Delta_bulk large Pr}), and (\ref{eq: Delta_bulk small Pr}), the resulting relation can be written as
\begin{eqnarray} \label{eq:analytic Nu-Ra-Pr}
	\frac{1}{2}
	&\approx&
	4.506\, \Nu^{3/4}\Ra^{-1/4}\Pran^{1/2}
	\left[1+0.7535\left(\frac{\Pran}{\Pran_T}\right)^{0.9338} \right]^{-1 / (0.9338 p^{\prime})}
	\nonumber \\[3pt]
	&& \mbox{} +  
	2.532\, {C_{\omega b}^{\prime}}^{-1/2} 
	\Nu^{3/4}\Ra^{-1/4}\Pran_T^{1/2} 
	f \!\left(\frac{\Pran}{\Pran_T}\right),
\end{eqnarray}
with
\begin{equation}
	C_{\omega b}^{\prime}
	=\frac{\beta_0}{\alpha\beta^{\ast}}-C_{\omega b}>0,
\end{equation}
and
\begin{equation}
	f(s)=
	\left\{
	\begin{array}{ll}
		\frac{4}{3}s^{-1/4}-\frac{1}{3}s^{1/2}, & 0<s<1, \\[6pt]
		s^{-1/2}, & s\ge 1 .
	\end{array}
	\right.
\end{equation}
The bridging function $f(s)$ is continuous and differentiable at $s=1$, ensuring a smooth crossover between the two regimes.

From Eq.~(\ref{eq:analytic Nu-Ra-Pr}), the asymptotic dependence of $\Nu$ on $\Ra$ and $\Pran$ follows as
\begin{equation}\label{eq: analytic Nu-Ra-Pr asymptotic}
	\Nu\sim 
	\left\{
	\begin{array}{ll}
		\Ra^{1/3}\Pran^{1/3}, & \Pran\ll 1, \\[4pt]
		\Ra^{1/3}\Pran^{-2/3+4/(3p^{\prime})}, & \Pran\gg 1 .
	\end{array}
	\right.
\end{equation}
For $\Pran\gg 1$, the exponent of $\Pran$ evaluates to $-0.4154$, using $p^{\prime}=5.307$.

The resulting $\Nu$ dependence on $\Ra$ and $\Pran$ predicted by the standard model exhibits a quantitative deviation from experimental observations.  
A detailed comparison with experiments, together with a new correction proposed to address this discrepancy, is presented in \S\ref{sec:4}.

\subsection{Validation with simulation}\label{sec:3.6}

The analytical solution derived in \S\ref{sec:3.1}--\ref{sec:3.5} is evaluated against statistically one-dimensional simulations. 
All simulations are performed for $C_{\omega b}=1$, a baseline value commonly adopted in the literature and used here to establish reference behavior prior to calibration. 
The governing equations are solved using the steady-state solver \texttt{buoyantSimpleFoam} of the open-source package OpenFOAM \citep{openfoam2025}. 
Because the $k$--$\omega$ formulation in OpenFOAM differs from that of \citet{wilcox2006turbulence}, the accurate form is implemented following the recommendations of \citet{NASA_TurbModels} and \citet{gomez2014accuracy}, with buoyant production included explicitly. 
All analytical results are determined solely by the prescribed $\Ra$ and $\Pran$, with $\Nu$ from Eq.~(\ref{eq:analytic Nu-Ra-Pr}) providing the basis for all dependent quantities.

\begin{figure}
	\centering
	\begin{subfigure}{0.48\textwidth}
		\includegraphics[scale=1.0]{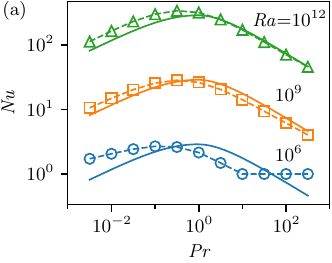}
	\end{subfigure}
	\begin{subfigure}{0.48\textwidth}
		\includegraphics[scale=1.0]{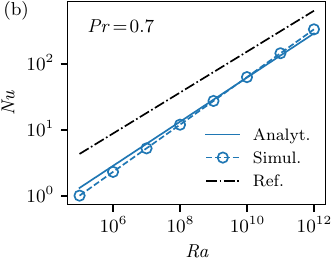}
	\end{subfigure}	
	\caption{
		One-dimensional Rayleigh--B\'enard convection predictions of the $k$--$\omega$ model with $C_{\omega b}=1$.  
		(a) Nusselt number as a function of the Prandtl number. Circles, squares, and triangles with dashed lines represent simulation results for $\Ra = 10^{6}$, $10^{9}$, and $10^{12}$, respectively, while solid lines denote the analytical prediction.  
		(b) Rayleigh-number dependence at $\Pran = 0.7$. The simulation results (circles) and analytical prediction (solid line) are compared with the experimental correlation of \citet{niemela2000turbulent} (dash-dotted line).
	}
	\label{fig:2}
\end{figure}

Figure~\ref{fig:2}(a) shows that the analytical prediction of $\Nu$ as a function of $\Ra$ and $\Pran$ agrees closely with the simulation results for $\Ra = 10^{9}$ and $10^{12}$ over the range $10^{-2.5} \le \Pran \le 10^{2.5}$. 
At $\Ra = 10^{6}$, the deviation arises because $\Nu$ approaches unity, rendering the thin-boundary-layer assumption invalid. 
Figure~\ref{fig:2}(b) presents the overall magnitude of the $\Nu$ dependence on $\Ra$ at fixed $\Pran = 0.7$ by comparison with the experimental correlation $\Nu = 0.124 \Ra^{0.309}$ of \citet{niemela2000turbulent}. 
With the conventional setting $C_{\omega b}=1$, the simulation results follow the power law $\Nu = 0.0163\Ra^{0.359}$, leading to an underprediction of approximately 50--70\% relative to the expected values. 
Overall, the analytical prediction agrees closely with the one-dimensional simulation results.

\begin{figure}
	\centering
	\begin{subfigure}{0.48\textwidth}
		\includegraphics[scale=1.0]{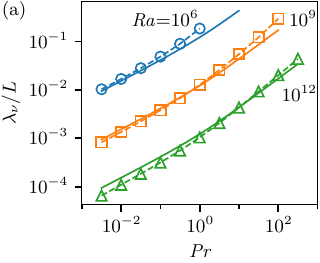}
	\end{subfigure}
	\begin{subfigure}{0.48\textwidth}
		\includegraphics[scale=1.0]{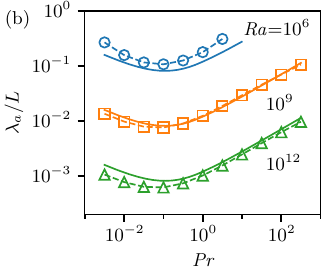}
	\end{subfigure}	
	\begin{subfigure}{0.48\textwidth}
		\includegraphics[scale=1.0]{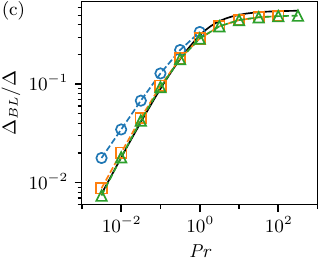}
	\end{subfigure}	
	\begin{subfigure}{0.48\textwidth}
		\includegraphics[scale=1.0]{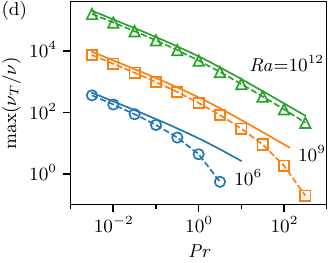}
	\end{subfigure}	
	\caption{
		One-dimensional Rayleigh--B\'enard convection predictions of the $k$--$\omega$ model with $C_{\omega b}=1$.  
		(a) Kinetic and (b) thermal boundary-layer thicknesses, (c) the temperature difference across the kinetic boundary layer, and (d) the maximum turbulent viscosity at the domain center as functions of the Prandtl number.  
		Circles, squares, and triangles with dashed lines denote simulation results for $\Ra = 10^{6}$, $10^{9}$, and $10^{12}$, respectively.  
		Solid lines represent the analytical relations.
	}
	\label{fig:3}
\end{figure}

Comparisons for additional thermal and turbulent quantities are presented in Fig.~\ref{fig:3}. 
The analytical expressions for $\lambda_\nu$ (Fig.~\ref{fig:3}a; Eq.~\ref{eq: kinetic BL}) and $\lambda_a$ (Fig.~\ref{fig:3}b; Eqs.~\ref{eq: thermal BL high Pr} and \ref{eq: thermal BL low Pr}) agree well with the simulations across the entire range investigated. 
The analytical near-wall temperature $\Delta_{BL}/\Delta$ (Eq.~\ref{eq: Delta_BL}) depends solely on $\Pran$ once $\Nu \sim \Ra^{1/3}$ is substituted, and Fig.~\ref{fig:3}(c) confirms that this relation remains valid and becomes increasingly accurate as $\Ra$ increases. 
The predicted maximum of $\nu_T/\nu$, shown in Fig.~\ref{fig:3}(d) and located at $z=L/2$, is also consistent with the simulations except when $\nu_T$ is not sufficiently larger than $\nu$. 
The analytical result for $\nu_T/\nu$ is obtained from Eqs.~(\ref{eq: k to omega at bulk}), (\ref{eq:tau definition}), and (\ref{eq: tau1}).

\begin{figure}
	\centering
	\begin{subfigure}{0.48\textwidth}
		\includegraphics[scale=1.0]{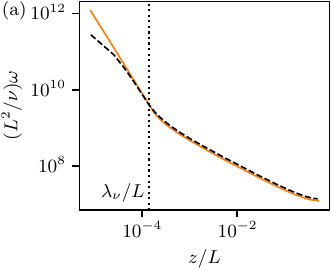}
	\end{subfigure}
	\begin{subfigure}{0.48\textwidth}
		\includegraphics[scale=1.0]{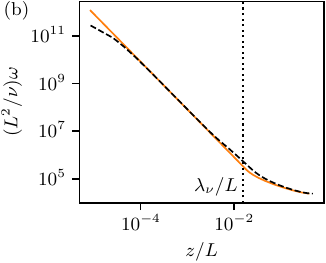}
	\end{subfigure}	
	\caption{
		One-dimensional Rayleigh--B\'enard convection predictions of the $k$--$\omega$ model with $C_{\omega b}=1$.  
		The distribution of $\omega$ at $\Ra = 10^{12}$ is shown for (a) $\Pran = 10^{-2}$ and (b) $\Pran = 10^{2}$.  
		Dashed lines denote simulation results, and solid lines denote the analytical relations.  
		Vertical dotted lines mark the analytical kinetic boundary-layer thicknesses.
	}
	\label{fig:4}
\end{figure}

The $\omega$ distributions at $\Ra = 10^{12}$ (Fig.~\ref{fig:4}) are well reproduced by the analytical forms, exhibiting the expected change in slope at $z=\lambda_\nu$ (\S\ref{sec:3.1}) for both small ($\Pran = 10^{-2}$; Fig.~\ref{fig:4}a) and large $\Pran$ ($\Pran = 10^{2}$; Fig.~\ref{fig:4}b).  
Minor discrepancies at the smallest $z$ arise because the singular distribution $\omega = 6\nu/(\beta_0 z^2)$ is implemented numerically as the fixed value $\omega = 6\nu/(\beta_0 n^2)$ at the wall, where $n$ denotes the distance to the first cell center (Eq.~\ref{eq: no slip BC}).  

In summary, the analytical framework reproduces the simulation results with high accuracy, establishing a consistent foundation for the buoyancy corrections developed in \S\ref{sec:4}.
\section{Formulation and calibration of buoyant corrections}\label{sec:4}

The analytical relation for $\Nu$ (Eq.~\ref{eq:analytic Nu-Ra-Pr}) depends not only on the control parameters $\Ra$ and $\Pran$, but also on the model constants. 
If $\Pran_T$ and $C_{\omega b}$ are treated as fixed, the model cannot modify the asymptotic $\Nu$--$\Ra$--$\Pran$ relations given by Eq.~(\ref{eq: analytic Nu-Ra-Pr asymptotic}). 
To enable calibration, these quantities must therefore be expressed as dimensionless functions, analogous to the auxiliary functions employed in standard $k$--$\omega$ formulations.

The calibration draws upon complementary datasets that collectively span wide ranges of $\Ra$ and $\Pran$.  
The experiments of \citet{niemela2000turbulent} established $\Nu \sim \Ra^{0.31}$ over $10^{6} \le \Ra \le 10^{17}$ for $\Pran \approx 0.7$, providing a robust benchmark for $\Ra$-scaling.  
Three-dimensional DNS studies further clarified the $\Pran$ dependence of heat transfer: $\Nu \sim \Pran^{0.14}$ at $\Ra=6\times10^{5}$ for $0.022\le\Pran\le0.7$ by \citet{verzicco1999prandtl}, and $\Nu \sim \Pran^{0.18}$ at $\Ra=10^{6}$ for $0.001\le\Pran\le1$ by \citet{pandey2022convective}.  
At higher $\Pran$, the simulations of \citet{li2021effects} exhibited an almost $\Pran$-independent behavior, $\Nu\sim\Pran^{-0.02}$, for $11.7\le\Pran\le650.7$ and $6.0\times10^{8}\le\Ra\le3.0\times10^{10}$.  
These findings are broadly consistent with the theoretical prediction of \citet{grossmann2000scaling}, which yields $\Nu\sim\Pran^{1/8}$ for $\Pran\ll1$ and $\Nu\sim\Pran^{0}$ for $\Pran\gg1$ at moderate $\Ra$.  
Taken together, these datasets delineate the relevant parameter space for calibration.

From a strict standpoint, these scaling laws depend on the ranges of $\Ra$ and $\Pran$, as well as on geometry, aspect ratio, wall roughness, and internal flow structure, and therefore cannot be represented by a single universal relation.  
Within the Reynolds-averaged framework, however, it is neither practical nor necessary to account for such dependencies explicitly.  
In engineering applications, where RANS models are primarily employed, the principal objective is to reproduce the correct mean Nusselt number rather than the detailed flow organization.  
As noted by \citet{ahlers2009heat}, parameters such as aspect ratio and large-scale circulation exert only a weak influence on the mean $\Nu$.  
Consequently, the adopted scaling laws are considered sufficient for calibration over the practically relevant range $10^{-3} \le \Pran \le 10^{3}$ and $\Ra \le 10^{10}$.

The standard model exhibits its largest discrepancy from experiments in the $\Pran>O(1)$ regime, where the predicted $\Nu\sim\Pran^{-0.4154}$ at fixed $\Ra$ contrasts with the nearly $\Pran$-independent behavior observed in DNS and experiments.  
For $\Pran<O(1)$, the prediction $\Nu\sim\Pran^{1/3}$ moderately departs from the experimental scaling $\Nu\sim\Pran^{0.14}$ to $\Pran^{0.18}$.  
In terms of $\Ra$ scaling, the observed $\Nu\sim\Ra^{0.31}$ is marginally lower than the model prediction $\Nu\sim\Ra^{1/3}$; this difference is sufficiently small that no correction is introduced.  
Overall, the standard model exaggerates the $\Pran$-dependence, producing exponents of larger magnitude in both limits, which motivates the introduction of the buoyant corrections developed in this study.

\subsection{Definition of the correction parameters}\label{sec:4.1}

The first step is to identify the parameters available for constructing the correction term.  
In the present formulation, the locally definable independent variables (Galilean invariants) reduce to six:
\[
k, \quad
\omega, \quad
\nu, \quad
a, \quad
gb, \quad
\frac{\mathrm{d}T}{\mathrm{d}z},
\]
where $g$ and $b$ appear only as the product $gb$ because gravity influences the flow exclusively through buoyancy.

These six dimensional variables, expressed in terms of temperature, time, and length scales, can be reduced to three nondimensional parameters:
\begin{equation}
	\Rey_T = \frac{\nu_T}{\nu}, \quad
	\Pran, \quad
	\frac{\mathcal{P}_b}{\varepsilon},
\end{equation}
where $\Rey_T$ is the turbulent Reynolds number, defined consistently with \citet{wilcox2006turbulence}, and the final term represents the ratio of buoyant production to the dissipation rate of $k$, with $\varepsilon=\beta^{\ast}k\omega$.

In the near-wall region ($z\le\lambda_\nu$), the analytical distributions of $\Rey_T$ and $\mathcal{P}_b/\varepsilon$ are
\begin{equation} \label{eq:Reyt}
	\Rey_T = \left( \frac{z}{\lambda_\nu} \right)^{p+2},
\end{equation}
\begin{equation}
	\frac{\mathcal{P}_b}{\varepsilon}
	= \left( \frac{z}{\lambda_\nu} \right)^{2-p}
	\frac{a_T/a}{1+a_T/a}
	= \frac{\Pran}{\Pran_T}
	\left( \frac{z}{\lambda_\nu} \right)^{4}
	\left[ 1 + \frac{\Pran}{\Pran_T}
	\left( \frac{z}{\lambda_\nu} \right)^{p+2} \right]^{-1},
\end{equation}
with $p=3.307$.  
For $\Pran>\Pran_T$, $\mathcal{P}_b/\varepsilon$ attains a maximum proportional to 
$(\Pran/\Pran_T)^{(p-2)/(p+2)}$, located at $z<\lambda_\nu$.

The turbulent Reynolds number reaches its maximum at $z=L/2$:
\begin{eqnarray}
	\Rey_T 
	&=& \frac{gbq}{\beta^{\ast}\nu}
	\left( \frac{L^2\tau_c}{\nu} \right)^{2} \nonumber \\[4pt]
	&=& \frac{1}{\beta^{\ast}}
	\left[
	\frac{2}{\pi^2}
	\frac{\beta^{\ast}}{\sigma}
	\left( \beta_0 - \alpha\beta^{\ast}C_{\omega b} \right)
	\right]^{2/3}
	\Ra^{1/3}\Nu^{1/3}\Pran^{-2/3}.
\end{eqnarray}

Under the present statistically one-dimensional Rayleigh--B\'enard configuration, the bulk value of $\mathcal{P}_b/\varepsilon$ is nearly uniform and close to unity.  
By contrast, when shear production is present, dissipation $\varepsilon$ balances the total production $(\mathcal{P}+\mathcal{P}_b)$, so that $\mathcal{P}_b/\varepsilon$ need not remain equal to unity.

In summary, any nondimensional buoyant correction function can be expressed in terms of the three parameters defined here, with its spatial variation inferred \textit{a priori} from the present analytical framework.

\subsection{Correction of $\Pran$ dependence}\label{sec:4.2}

The standard model shows its largest discrepancy from experiments in the $\Nu$--$\Pran$ relation at fixed $\Ra$, since the $\Pran$ dependence of buoyancy effects has not been explicitly addressed in previous formulations.  
This subsection corrects that dependence within the analytical framework established in \S\ref{sec:3}.

Including the model constants associated with buoyant thermal convection, the asymptotic form of Eq.~(\ref{eq:analytic Nu-Ra-Pr}) becomes
\begin{equation}\label{eq: analytic Nu-Ra-Pr constant}
	\Nu\sim \left\{
	\begin{array}{ll}
		\Ra^{1/3} 
		\left( 
		\Pran^{1/2} 
		+ {C_{\omega b}^{\prime}}^{-1/2} 
		\Pran_T^{3/4}
		\Pran^{-1/4}
		\right)^{-4/3},
		& \Pran\ll \Pran_T \\[4pt]
		\Ra^{1/3} 
		\left( 
		\Pran_T^{0.1884}
		\Pran^{0.3116}
		+  {C_{\omega b}^{\prime}}^{-1/2} 
		\Pran_T
		\Pran^{-1/2}
		\right)^{-4/3},
		& \Pran\gg \Pran_T .
	\end{array}
	\right.
\end{equation}
Equation~(\ref{eq: analytic Nu-Ra-Pr constant}) indicates that the $\Pran$ dependence of $\Nu$ can be controlled through $C_{\omega b}^{\prime}$ and $\Pran_T$.
The terms ${C_{\omega b}^{\prime}}^{-1/2}\Pran_T^{3/4}\Pran^{-1/4}$ and $\Pran_T^{0.1884}\Pran^{0.3116}$ dominate in the low- and high-$\Pran$ limits, respectively, corresponding to the contributions of $\Delta_{bulk}$ and $\Delta_{BL}$ identified in \S\ref{sec:3.5}.  
Accordingly, two complementary corrections emerge naturally:
\begin{enumerate}
	\item a bulk correction to $C_{\omega b}^{\prime}$ for $\Pran \ll 1$, and
	\item a near-wall modification of $a_T$ for $\Pran \gg 1$.
\end{enumerate}

For correction (1), the uncorrected model predicts $\Nu \sim \Pran^{1/3}$ in the low-$\Pran$ regime; adopting a smaller asymptotic scaling exponent is therefore effective for reproducing the observed behavior $\Nu \sim \Pran^{0.14}$–$\Pran^{0.18}$ over $0.001 \le \Pran \le 1$.
Motivated by this consideration, the model targets the theoretical scaling $\Nu \sim \Pran^{1/8}$ for $\Pran \ll 1$ proposed by \citet{grossmann2000scaling}. 
To meet this target, the model must satisfy
\begin{equation}
	\left( {C_{\omega b}^{\prime}}^{-1/2} 
	\Pran_T^{3/4}
	\Pran^{-1/4}
	\right)^{-4/3} \sim \Pran^{1/8},
\end{equation}
which implies
\begin{equation}
	C_{\omega b}^{\prime} \sim \Pran^{-5/16} 
	\quad \mathrm{for} \quad \Pran \ll 1.
\end{equation}
Here, the standard value $\Pran_T = 0.89$ is retained to avoid introducing unnecessary modifications to the model.

The correction to $C_{\omega b}^{\prime}$ is applied uniformly throughout the domain and is expressed as
\begin{equation}\label{eq: Cwb correction}
	C_{\omega b}^{\prime}
	= c_1 + c_2 \Pran^{-5/16}.
\end{equation}
Here, the coefficients $c_1$ and $c_2$ determine both the Prandtl number at which the model enters the low-$\Pran$ regime and the corresponding level of $\Nu$ in that regime.

The discussion now turns to correction (2): a near-wall modification of $a_T$ for the large-$\Pran$ regime.
To recover $\Nu \sim \Pran^{0}$ for $\Pran \gg 1$, so that $\Delta_{BL}/\Delta \sim \Pran^{0}$, the near-wall integral term in Eq.~(\ref{eq: Delta_BL original}) must satisfy
\begin{equation}
	\int_{0}^{1} \frac{1}{1+a_T/a} \mathrm{d}z^{\prime} 
	\sim \left\{
	\begin{array}{ll}
		1, & \Pran\ll 1 \\[2pt]
		\Pran^{-1/2}, & \Pran\gg1,
	\end{array}
	\right.
\end{equation}
where $z^{\prime}=z/\lambda_\nu$.  
The modeling objective is thus to design the near-wall $a_T$ distribution so that this integral reproduces the target scaling.  
Following the approach in \S\ref{sec:3.4.1}, the wall integral is approximated as
\begin{equation}\label{eq: Pr Correction integral target}
	\int_{0}^{1} \frac{1}{1+a_T/a} \mathrm{d} z^{\prime} 
	\approx \left( 1 + d_1 \Pran^{d_2/2} \right)^{-1/d_2},
\end{equation}
Analogously to $c_1$ and $c_2$, the coefficients $d_1$ and $d_2$ determine both the Prandtl-number threshold for entering the high-$\Pran$ regime and the corresponding magnitude of $\Nu$.

\begin{figure}
	\centering
	\begin{subfigure}{0.56\textwidth}
		\includegraphics[scale=1.0]{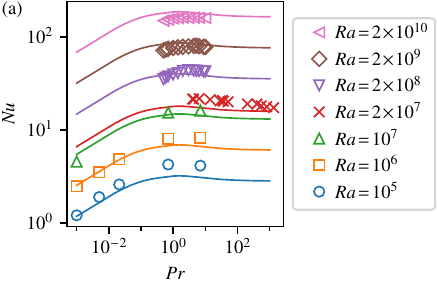}
	\end{subfigure}
	\begin{subfigure}{0.35\textwidth}
		\includegraphics[scale=1.0]{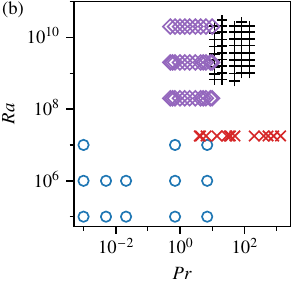}
	\end{subfigure}
	\caption{
		(a) Dependence of $\Nu$ on $\Pran$ using the applied corrections (Eqs.~\ref{eq: Cwb correction} and \ref{eq: Pr Correction integral target}) in one-dimensional Rayleigh--B\'enard convection. 
		Markers denote reference data, and the overlapping solid lines indicate the corrected analytical relations.
		(b) Reference datasets used for the correction, shown in the $\Ra$--$\Pran$ space: circles, crosses, diamonds, and pluses denote the datasets of \citet{pandey2022convective}, \citet{xia2002heat}, \citet{stevens2011prandtl}, and \citet{li2021effects}, respectively.
	}
	\label{fig:5}
\end{figure}

The coefficients $c_1$, $c_2$, $d_1$, and $d_2$ are determined from reference data.
Datasets are employed to span wide ranges of $\Ra$ and $\Pran$:  
$10^{-3} \le \Pran \le 7$ and $10^{5} \le \Ra \le 10^{7}$ from \citet{pandey2022convective};  
$4 \le \Pran \le 1.3\times10^{3}$ at $\Ra = 1.78\times10^{7}$ from \citet{xia2002heat};  
$0.5 \le \Pran \le 10$ and $2\times10^{8} \le \Ra \le 2\times10^{10}$ from \citet{stevens2011prandtl}; and  
$11.7 \le \Pran \le 145.7$ and $6\times10^{8} \le \Ra \le 3\times10^{10}$ from \citet{li2021effects}.  
The constants $c_1$, $c_2$, $d_1$, and $d_2$ are obtained by minimizing the global least-squares error between the analytical $\Nu$ and the data using the differential-evolution algorithm \citep{storn1997differential}, implemented as \texttt{scipy.optimize.differential\_evolution} in Python~\citep{ralf_gommers_2025_17101542}.  
The optimal values are $c_1=2.488$, $c_2=0.2988$, $d_1=1.125$, and $d_2=1.173$.  
Figure~\ref{fig:5}(a) demonstrates that the corrected analytical relation reproduces the observed $\Pran$ dependence across both low- and high-$\Pran$ regimes.  
Figure~\ref{fig:5}(b) shows all employed data points in the $\Ra$--$\Pran$ space.

To realise the target integral relation defined in Eq.~(\ref{eq: Pr Correction integral target}), 
a near-wall correction function $\psi$ is introduced to extend the conventional formulation $a_T = \nu_T / \Pran_T$:
\begin{equation}\label{eq: Pr correction g}
	\frac{a_T}{a} = \frac{\nu_T}{ \Pran_T a} 
	+ \psi\!\left(\Rey_T , \Pran, \frac{\mathcal{P}_b}{\varepsilon}\right).
\end{equation}
The dimensionless function $\psi$ must satisfy the following conditions:  
(i) $\psi = 0$ at the wall, because $a_T$ must vanish at the wall.  
(ii) $\psi = 0$ for $\mathcal{P}_b / \varepsilon = 0$, ensuring that the correction leaves the standard formulation unchanged in the absence of buoyancy.  
(iii) $\psi \ll \nu_T / (\Pran_T a)$ for $z \gg \lambda_\nu$, indicating that the correction acts only as a near-wall modification.  
(iv) $\psi \gg \nu_T / (\Pran_T a)$ for $z \ll \lambda_\nu$, thereby enhancing the near-wall turbulent heat flux.
This increase compensates for the underprediction of $\Nu$ at high $\Pran$ in the uncorrected model.
In detail, the integral of the corrected $a_T$ must satisfy Eq.~(\ref{eq: Pr Correction integral target}) with $d_1=1.125$ and $d_2=1.173$.

A convenient functional form satisfying these conditions is
\begin{equation}
	\psi = e_0  
	\Pran^{e_1}
	\left( \frac{\mathcal{P}_b}{\varepsilon} \right)^{e_2}
	\Rey_T^{e_3}.
\end{equation}
The corresponding near-wall distribution of $a_T/a$ for $z \le \lambda_\nu$ becomes
\begin{equation}\label{eq: aT wall restriction}
	\frac{a_T}{a}
	= \frac{\Pran}{\Pran_T}\!\left(\!\frac{z}{\lambda_\nu}\!\right)^{p+2}
	+ e_0 \Pran^{e_1}
	\!\left(\!\frac{z}{\lambda_\nu}\!\right)^{(2-p)e_2+(p+2)e_3}
	\!\left(\!\frac{a_T/a}{1+a_T/a}\!\right)^{e_2}.
\end{equation}
The conditions on $\psi$ imply
(i) $0 < (2-p)e_2 + (p+2)e_3$;
(ii) $0 < e_2$;
(iii) $e_3 < 1$; and
(iv) $e_2 < 1$ and $\left[(2-p)e_2 + (p+2)e_3\right]/(1 - e_2) < p + 2$.
Combining (i) and (ii) yields $0 < e_3$.  
For condition (iv), the quantity $\left[(2-p)e_2 + (p+2)e_3\right]/(1 - e_2)$ represents the exponent governing the near-wall power-law asymptotic scaling of $a_T/a$ with respect to $z$ as $z \rightarrow 0$, valid when $e_2 < 1$.

Since only $e_2$ and $e_3$ are associated with the necessary condition, they are considered first.
Values of $e_2 = e_3 = 1/2$ are adopted to ensure sufficient robustness, as they lie near the middle of the admissible ranges and satisfy all constraints.
The remaining parameters $e_0$ and $e_1$ are optimised to reproduce the integral relation~(\ref{eq: Pr Correction integral target}) for $10^{2} \le \Pran \le 10^{3}$ using the same differential-evolution method, yielding $e_0=7.141$ and $e_1=0.8974$.

\begin{figure}
	\centering
	\begin{subfigure}{0.48\textwidth}
		\includegraphics[scale=1.0]{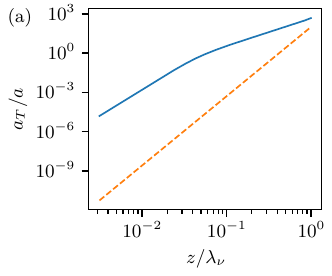}
	\end{subfigure}
	\begin{subfigure}{0.48\textwidth}
		\includegraphics[scale=1.0]{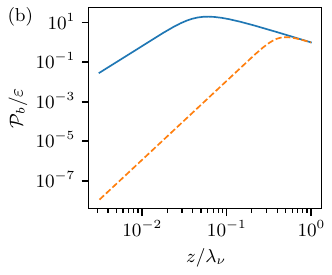}
	\end{subfigure}
	\caption{
		Analytical distributions at $\Pran=10^2$:  
		(a) the turbulent thermal diffusivity and (b) the ratio of buoyant production to dissipation.  
		Dashed and solid lines represent the original and corrected models, respectively.
	}
	\label{fig:6}
\end{figure}

As shown in figure~\ref{fig:6}(a), the correction increases $a_T$ near the wall while leaving the bulk region almost unchanged.  
For $e_2 = e_3 = 1/2$, the asymptotic near-wall scaling (as $z \to 0$) of $a_T/a$ changes from $z^{p+2} = z^{5.307}$ to $z^{4}$.  
The corresponding distribution of $\mathcal{P}_b / \varepsilon$, shown in figure~\ref{fig:6}(b), changes from $z^{4}$ to $z^{6-p} = z^{2.693}$.  
These modifications strengthen the turbulent heat flux in the near-wall region, thereby compensating for the underprediction of $\Nu$ at high $\Pran$.  
As a result, the corrected model increases $\Nu$ for $\Pran > 1$ while maintaining consistency with the standard framework.

In summary, the proposed corrections are
\begin{equation}\label{eq: Cwb correction Pr}
	C_{\omega b} =  -0.9752 
	- 0.2988 \Pran^{-5/16} \quad \mathrm{for} \quad \mathcal{P}_b>0,
\end{equation}
\begin{equation}\label{eq: aT correction final}
	\frac{a_T}{a} = 
	\frac{\Pran}{\Pran_T} \frac{\nu_T}{\nu}
	+ 7.141 \Pran^{0.8974}
	\!\left[
	\min \!\left(
	\max\!\left(\frac{\mathcal{P}_b}{\varepsilon},0\right)
	, C_{lim,b}
	\right)
	\!\right]^{1/2}
	\!\left(\frac{\nu_T}{\nu}\right)^{1/2}.
\end{equation}
An upper limiter $C_{lim,b}$ is introduced solely to ensure numerical robustness.  
During the early stages of a calculation, certain poorly scaled initial conditions
(such as large $k$ combined with very small $\omega$) can lead to transiently large
values of $\mathcal{P}_b/\varepsilon$, sometimes far exceeding unity. 
Such excursions amplify $a_T$ locally, producing steep temperature gradients through
the heat--balance relation, which then feed back into $\mathcal{P}_b$ and may cause
instabilities.  
To prevent this behavior, the value is capped at $C_{lim,b}=100$.  
This threshold is far above the converged values of $\mathcal{P}_b/\varepsilon$ and
therefore does not influence the final solution, while providing a safeguard
against nonphysical transients during the convergence process.

\begin{figure}
	\centering
	\begin{subfigure}{0.48\textwidth}
		\includegraphics[scale=1.0]{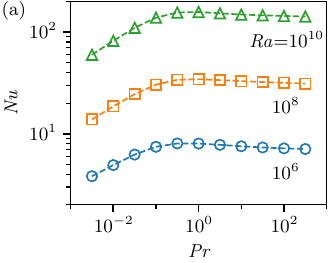}
	\end{subfigure}
	\begin{subfigure}{0.48\textwidth}
		\includegraphics[scale=1.0]{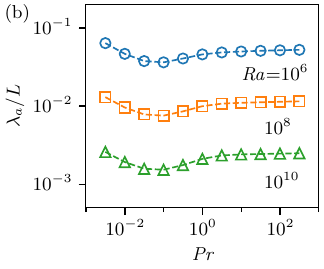}
	\end{subfigure}
	\caption{
		Simulation results for one-dimensional Rayleigh--B\'enard convection with the $\Pran$ correction: (a) Nusselt number and (b) thermal boundary-layer thickness as functions of $\Pran$.
		The solid lines in panel (a) represent the analytical solution of the corrected model.
	}
	\label{fig:7}
\end{figure}

Figure~\ref{fig:7} presents the one-dimensional simulation results obtained with the $\Pran$ correction, using the same numerical method described in \S\ref{sec:3.6}.  
As shown in figure~\ref{fig:7}(a), the corrected $a_T$ (Eq.~\ref{eq: aT correction final}) successfully reproduces the $\Nu \sim \Pran^{0}$ scaling for $\Pran > 1$.  
For $\Ra = 10^{6}$, the simulations yield $\Nu = 13.1 \Pran^{0.213}$ over $10^{-2.5} \le \Pran \le 10^{-1.5}$, consistent with the reference data of \citet{pandey2022convective}, which reported $\Nu = 11.3 \Pran^{0.220}$ for $0.001 \le \Pran \le 0.021$.  
These results indicate that the correction to $C_{\omega b}$ effectively reproduces the $\Nu$ dependence in the low-$\Pran$ regime.  
As illustrated in figure~\ref{fig:7}(b), the correction also improves the prediction of the thermal boundary-layer thickness.  
The DNS of \citet{stevens2011prandtl} show that the thermal boundary-layer thickness remains nearly independent of $\Pran$ for $\Pran>1$.
The corrected model reproduces this behavior, whereas the uncorrected model erroneously predicts $\lambda_a/L \sim \Pran^{0.4041}$ (see figure~\ref{fig:3}b).
Overall, the proposed buoyancy correction substantially enhances the predictive performance of the standard $k$--$\omega$ model while preserving internal consistency and numerical robustness.

\subsection{$C_{\omega b}$ for stable stratification}\label{sec:4.3}

After calibrating $C_{\omega b}$ for unstable stratification ($\mathcal{P}_b > 0$), the analysis is extended to the stably stratified regime ($\mathcal{P}_b < 0$) to complete the formulation of the $\omega$ equation.
As summarized in \S\ref{sec:2.2}, \citet{rodi1987examples} recommended the empirical setting $C_{\varepsilon b}=0$ for $\mathcal{P}_b<0$, an approach that has been widely adopted.  
Through the correspondence between the $k$--$\varepsilon$ and $k$--$\omega$ formulations (Eq.~\ref{eq: Cwb Ceb}), this prescription corresponds to $C_{\omega b}\approx -2$.
As an alternative to empirical calibration, the present study examines an analytical procedure for determining $C_{\omega b}$ for $\mathcal{P}_b<0$.

This analytical approach is guided by the canonical case of homogeneous stably stratified shear flow, which provides a well-defined stability threshold for model consistency.
As discussed in \S\ref{sec:1}, models for $\mathcal{P}_b<0$ are typically constrained by the behavior of stably stratified shear flows, where linear stability analyses predict flow stabilization once the flux Richardson number $\Rif=-\mathcal{P}_b/\mathcal{P}$ exceeds approximately $0.25$, as reported by \citet{gibson1978ground,durbin2011statistical}.  
Following this benchmark, \citet{burchard1995performance,burchard2007applied} analyzed the $k$--$\varepsilon$ model from a dynamical-systems perspective to determine the appropriate form of $C_{\varepsilon b}$ for stably stratified shear flows.
This framework treats the $k$--$\varepsilon$ model as a set of ordinary differential equations for $k(t)$ and $\varepsilon(t)$, whose trajectories in the $(k,\varepsilon)$ phase space are analyzed to determine the stabilization conditions.  
For further discussion on dynamical-systems analyses of RANS models, see \citet{pettersson2006behavior,rumsey2006arbitrary,rumsey2007apparent}.

In the same framework, the appropriate $C_{\omega b}$ for $\mathcal{P}_b<0$ can be determined analytically.  
The idealized configuration corresponds to a homogeneous stably stratified shear flow in which $k$ and $\omega$ vary only with time.  
Under these conditions, the production terms become
\[
\mathcal{P} = \nu_T S^2 = S^2 \frac{k}{\omega}, \qquad
\mathcal{P}_b = -g\beta \frac{\nu_T}{\Pran_T}\frac{\mathrm{d}T}{\mathrm{d}z}
= -\left(\frac{g\beta}{\Pran_T}\frac{\mathrm{d}T}{\mathrm{d}z}\right)\frac{k}{\omega}
= -\Rif\mathcal{P},
\]
where $S^2=S_{ij}S_{ij}$ and $\mathrm{d}T/\mathrm{d}z>0$, so that $\mathcal{P}>0$ and $\mathcal{P}_b<0$ with constant $S$ and $\Rif>0$.

Under this configuration, the $k$--$\omega$ equations reduce to
\begin{equation}
	\frac{\mathrm{d}k}{\mathrm{d}t}
	= k\!\left[\frac{(1-\Rif)S^2}{\omega}-\beta^{\ast}\omega\right],
\end{equation}
\begin{equation}
	\frac{\mathrm{d}\omega}{\mathrm{d}t}
	= \alpha S^2(1-\Rif C_{\omega b})-\beta_0\omega^2.
\end{equation}

Because $\omega$ evolves independently of $k$, its temporal behavior is examined first.  
By definition, $\omega(t)$ must remain positive for any $\Rif$, implying that the source term must stay non-negative even in the limit of large $\Rif$.
This constraint implies
\begin{equation}
	C_{\omega b}\le0 \quad \mathrm{for}\quad \mathcal{P}_b<0,
\end{equation}
as a positive $C_{\omega b}$ would drive $\mathrm{d}\omega/\mathrm{d}t<0$ near $\omega\to0$, yielding unphysical negative values of $\omega$.

For $C_{\omega b}\le0$, $\omega$ asymptotically approaches a steady state ($t\to\infty$) given by
\begin{equation}
	\omega_{\infty}
	=\left[\frac{\alpha S^2}{\beta_0}(1-\Rif C_{\omega b})\right]^{1/2},
\end{equation}
which is readily seen to be a stable equilibrium, as $\mathrm{d}\omega/\mathrm{d}t<0$ for $\omega>\omega_{\infty}$ and $\mathrm{d}\omega/\mathrm{d}t>0$ for $\omega<\omega_{\infty}$.

Substituting $\omega_{\infty}$ into the $k$ equation gives
\begin{equation}
	\frac{\mathrm{d}k}{\mathrm{d}t}
	= k\!\left[\frac{(1-\Rif)S^2}{\omega_{\infty}}-\beta^{\ast}\omega_{\infty}\right].
\end{equation}
For the turbulent kinetic energy to decay with time, the right-hand side must be negative:
\[
\frac{(1-\Rif)S^2}{\omega_{\infty}}-\beta^{\ast}\omega_{\infty}<0.
\]
Rearranging yields the stability condition
\begin{equation}
	\Rif>
	\left(1-\frac{\alpha\beta^{\ast}}{\beta_0}\right)
	\!\left(1-\frac{\alpha\beta^{\ast}}{\beta_0}C_{\omega b}\right)^{-1}.
\end{equation}
Hence, the value of $C_{\omega b}$ that reproduces the conventional stabilization threshold of $\Rif \ge 0.25$ is
\begin{equation}\label{eq: Cwb correction stably}
	C_{\omega b}
	=\frac{\beta_0}{\alpha\beta^{\ast}}
	-\frac{1}{0.25}
	\left(\frac{\beta_0}{\alpha\beta^{\ast}}-1\right)
	=-0.5385
	\quad \mathrm{for}\quad \mathcal{P}_b<0.
\end{equation}
By comparison, the earlier empirical recommendation $C_{\omega b}\approx-2$ would predict stabilization at $\Rif \ge 0.146$, lower than the benchmark.

Combining Eqs.~(\ref{eq: Cwb correction Pr}) and (\ref{eq: Cwb correction stably}) and substituting them into (\ref{eq: omega eq}) gives the complete $\omega$ equation:
\begin{eqnarray}\label{eq: Cwb correction final}
	\frac{\partial\omega}{\partial t}
	+U_j\frac{\partial\omega}{\partial x_j}
	&=&
	\alpha\frac{\omega}{k}
	\left[
	\mathcal{P}
	+C_{\omega b}^{+}\max(\mathcal{P}_b,0)
	+C_{\omega b}^{-}\min(\mathcal{P}_b,0)
	\right]
	\nonumber\\[3pt]
	&&{}-\beta_0f_{\beta}\omega^2
	+\frac{\sigma_d}{\omega}
	\frac{\partial k}{\partial x_j}
	\frac{\partial\omega}{\partial x_j}
	+\frac{\partial}{\partial x_j}
	\!\left[\!\left(\nu+\sigma\frac{k}{\omega}\right)
	\frac{\partial\omega}{\partial x_j}\!\right],
\end{eqnarray}
where the buoyancy coefficients are
\begin{equation}\label{eq: Cwb correction final2}
	C_{\omega b}^{+}=-0.9752 - 0.2988 \Pran^{-5/16}, \qquad 
	C_{\omega b}^{-}=-0.5385.
\end{equation}

As a further quantitative validation, a stably stratified plane channel flow is considered.  
The upper hot wall ($T=\Delta$ at $z=h$) and the lower cold wall ($T=0$ at $z=-h$) are both isothermal and no-slip.  
Gravity acts in the vertically downward ($-z$) direction, while the flow is driven horizontally.  
The dimensionless control parameters are the friction Reynolds number, $\Rey_\tau = U_\tau h/\nu$, where $U_\tau$ ($=\sqrt{\nu \mathrm{d}U/\mathrm{d}z}$ at the walls) is the friction velocity and $h$ is the channel half-height, and the friction Richardson number, $\mathit{Ri}_\tau = g b \Delta h / U_\tau^{2}$.  
The dimensionless output quantities are the Nusselt number, $\Nu = 2h q / (a \Delta)$, and the bulk Reynolds number, $\Rey_b = U_b h / \nu$, where $U_b$ is the bulk mean velocity.

The choices $C_{\omega b} = -0.5385$ and $C_{\omega b} = -2$ are tested under the same conditions as the DNS of \citet{garcia2011turbulence}.
As a baseline, the case without buoyancy ($\mathit{Ri}_\tau = 0$) is first examined using the standard $k$--$\omega$ model:  
(i) for $\Rey_\tau = 180$, DNS yields $\Rey_b = 2.82\times10^{3}$ and $\Nu = 6.03$, while RANS yields $\Rey_b = 2.75\times10^{3}$ and $\Nu = 6.36$;  
(ii) for $\Rey_\tau = 550$, DNS yields $\Rey_b = 1.01\times10^{4}$ and $\Nu = 16.44$, while RANS yields $\Rey_b = 1.03\times10^{4}$ and $\Nu = 17.0$.  
Here, $\Pran = 0.7$.  
Thus, for this reference case, the standard model predicts the mean heat transport within approximately 6\% relative error.

\begin{figure}[h]
	\centering
	\begin{subfigure}{0.48\textwidth}
		\centering
		\includegraphics[scale=1.0]{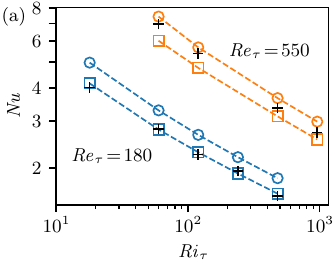}
	\end{subfigure}
	\begin{subfigure}{0.48\textwidth}
		\centering
		\includegraphics[scale=1.0]{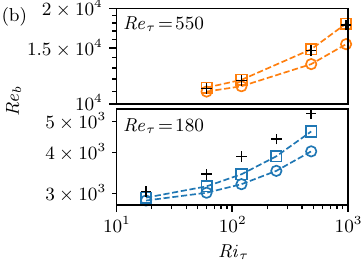}
	\end{subfigure}
	\caption{
		Stably stratified channel flow at $\Pran=0.7$.
		(a) Nusselt number and (b) bulk Reynolds number as functions of the friction Richardson number, at fixed friction Reynolds numbers of 180 (upper curves) and 550 (lower curves).
		Circles and squares denote the simulation results for $C_{\omega b}=-0.5385$ and $C_{\omega b}=-2$, respectively.
		Plus signs denote the DNS data of \citet{garcia2011turbulence}.
	}
	\label{fig:8}
\end{figure}

Figure~\ref{fig:8} shows the model predictions of (a) $\Nu$ and (b) $\Rey_b$ as functions of $\mathit{Ri}_\tau$ at fixed $\Rey_\tau = 180$ and $\Rey_\tau = 550$.
When $C_{\omega b}=-0.5385$ is used, the model predicts larger $\Nu$ and smaller $\Rey_b$ relative to $C_{\omega b}=-2$: this occurs because the source term in the $\omega$ equation becomes smaller, leading to smaller $\omega$ and therefore larger $\nu_T=k/\omega$.
Within the present parameter range, $C_{\omega b}=-2$ generally yields better agreement, but the trends indicate that the predictions for $C_{\omega b}=-0.5385$ become closer to DNS at higher $\Rey_\tau$.
For $\Rey_\tau=550$, the $\Nu$ prediction using $C_{\omega b}=-0.5385$ shows a slight improvement.
Overall, the differences in the dimensionless mean quantities associated with the two constants remain within about 20\%.

In summary, while the analytically derived value $C_{\omega b}=-0.5385$ produces a stability threshold that is internally consistent, the quantitative accuracy of both $C_{\omega b}=-2$ and $C_{\omega b}=-0.5385$ is configuration- and parameter-dependent. Even in a purely Reynolds-averaged sense, buoyancy damping affects only the gravity-aligned component of $k$, which induces strong anisotropy in the Reynolds stresses — a feature that is not naturally accommodated within the Boussinesq eddy-viscosity closure of the standard model. Higher-order treatments at the level of Reynolds-stress and heat-flux closures can be found in \citet{lazeroms2013explicit}, but a more comprehensive treatment of such issues lies beyond the present scope, which is limited to optimizing the model within the standard $k$--$\omega$ formulation.

\section{Model evaluation}\label{sec:5}

This section assesses the performance of the buoyancy-corrected $k$--$\omega$ model across a range of buoyancy-driven flows, extending beyond the statistically one-dimensional Rayleigh--B\'enard configuration.
The corrected formulation preserves the structure of the standard model and differs only through the buoyancy-related modifications introduced in Eqs.~(\ref{eq: aT correction final}, \ref{eq: Cwb correction final}, \ref{eq: Cwb correction final2}).
For evaluating the uncorrected standard model, the conventional choices $C_{\omega b}=1$ for $\mathcal{P}_b>0$ and $C_{\omega b}=-2$ for $\mathcal{P}_b<0$ are adopted.
The simulation procedure is identical to that described in \S\ref{sec:3.6}, and steady solutions are obtained by neglecting the temporal derivatives and iterating until convergence.

Previous RANS studies of buoyancy-driven turbulence have largely focused on a limited set of canonical configurations.  
Vertically heated natural convection in enclosed cavities has been the most widely explored \citep{ince1989computation, henkes1991natural, peng1999computation, dol2001computational, kenjerevs2005contribution, dehoux2017elliptic}, typically within the ranges $\Pran = 0.7$--$7$ and $10^{8} \le \Ra \le 10^{15}$.  
Rayleigh--B\'enard convection has also been examined \citep{kenjerevs1995prediction, choi2012turbulence}, although generally for $\Ra \le 10^{9}$ at fixed $\Pran = 0.7$.  
Vertical natural convection in channels \citep{dol1999dns, dehoux2017elliptic} has received comparatively less attention and is usually restricted to $\Ra = 10^{5}$--$10^{7}$ at $\Pran = 0.7$.

Despite spanning several orders of magnitude in Rayleigh number, these investigations cover only a small number of geometries and share common limitations.  
Systematic evaluation of Prandtl-number effects is rare, and many buoyancy-correction approaches rely on ad hoc functional forms whose validity has been demonstrated only within restricted flow regimes.  
As a result, the broader applicability of these corrections remains uncertain.

The present study examines a wider variety of buoyancy-driven flows:  
(i) two internally heated convection configurations;  
(ii) unstably stratified Couette flow (mixed convection);  
(iii) Rayleigh--B\'enard convection in a two-dimensional square cavity; and  
(iv) vertically heated natural convection in two-dimensional rectangular cavities with aspect ratios of 1 and 4.

Vertical channel flow is not included.  
Although buoyancy formally drives the flow in this configuration, the imposed temperature gradient is strictly perpendicular to gravity.
Under the gradient-diffusion hypothesis used for turbulent heat-flux modeling, this alignment yields $\mathcal{P}_b = 0$ throughout the domain, and the buoyancy-correction terms introduced here therefore have no effect.

\subsection{Internally heated convection in two configurations}\label{sec:5.1}

\begin{figure}[h]
	\centering
	\begin{subfigure}{0.3\textwidth}
		\centering
		\includegraphics[scale=1.0]{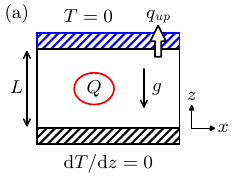}
	\end{subfigure}
	\begin{subfigure}{0.3\textwidth}
		\centering
		\includegraphics[scale=1.0]{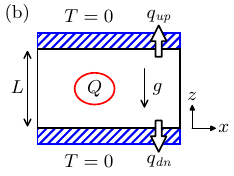}
	\end{subfigure}
	\caption{Internally heated convection setup for (a) top cooling and (b) top-and-bottom cooling condition.}
	\label{fig:9}
\end{figure}

Internally heated convection is used here to assess whether the correction derived for Rayleigh--B\'enard convection also applies to buoyancy-driven flows whose temperature distributions differ fundamentally from the Rayleigh--B\'enard configuration.  
A spatially uniform volumetric heat source $Q$ (defined in Eq.~\ref{eq: T eq}) is imposed throughout the statistically one--dimensional domain. Two configurations are considered, as illustrated in Figure~\ref{fig:9}:  

(a) Top cooling: the upper wall is isothermal and cooled, while the lower wall is adiabatic; that is, $T=0$ at $z=L$ and $\mathrm{d}T/\mathrm{d}z=0$ at $z=0$. In this case, the upward heat flux $q_{\mathit{up}}$ ($=-a \mathrm{d}T/\mathrm{d}z$ at $z=L$) satisfies $q_{\mathit{up}}=QL$.  

(b) Top-and-bottom cooling: both the upper and lower walls are isothermal at the same temperature, such that $T=0$ at $z=0$ and $z=L$. The upward and downward heat fluxes ($q_{\mathit{up}}$ and $q_{\mathit{dn}}$) satisfy the global balance $QL = q_{\mathit{up}} + q_{\mathit{dn}}$. 

For both configurations, the top and bottom walls satisfy no-slip boundary conditions.
To characterize the flow, the modified Rayleigh number is defined as $\Ra^{\prime}=g b Q L^5 / (\nu a^2)$. The dimensionless temperature is $T^{\ast}=aT/(L^2 Q)$, and either the domain--averaged temperature $T_{\mathit{avg}}^{\ast}$ or the maximum temperature $T_{\mathit{max}}^{\ast}$ is used for comparison against reference data. For the top--and--bottom cooling configuration (Figure~\ref{fig:9}(b)), an additional output parameter is introduced---the fraction of the total generated heat escaping through the lower wall, $F_{\mathit{dn}}=q_{\mathit{dn}}/(QL)$.

\begin{figure}[h]
	\centering
	\begin{subfigure}{0.48\textwidth}
		\centering
		\includegraphics[scale=1.0]{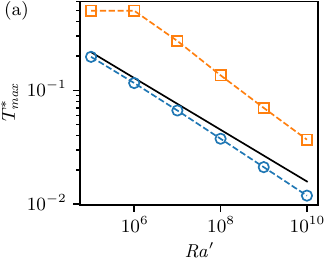}
	\end{subfigure}
	\begin{subfigure}{0.48\textwidth}
		\centering
		\includegraphics[scale=1.0]{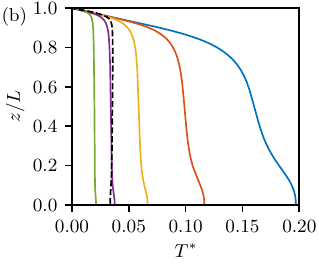}
	\end{subfigure}
	\caption{Internally heated convection with top cooling at $\Pran=6$.
		(a) Maximum dimensionless temperature as a function of the internal Rayleigh number.
		The solid line is the experimental correlation of \citet{kulacki1977steady}; 
		circles are the corrected $k$--$\omega$ model;  
		squares are the standard $k$--$\omega$ model with $C_{\omega b}=1$.
		(b) Dimensionless temperature profiles.  
		The solid lines correspond to the corrected $k$--$\omega$ model at $\Ra^{\prime}=10^{5}$, $10^{6}$, $10^{7}$, $10^{8}$ and $10^{9}$, ordered from right to left.  
		The dashed line shows the DNS results of \citet{nourgaliev1997investigation} at $\Ra=1.86\times10^{8}$.
	}
	\label{fig:10}
\end{figure}

Figure~\ref{fig:10} shows simulation results for the top-cooling configuration at $\Pran = 6$, with panel (a) presenting the maximum dimensionless temperature. 
The experimental correlation of \citet{kulacki1977steady} is $T_{\mathit{max}}^{\ast}=2.96 (\Ra^{\prime})^{-0.227}$ for $2.75\le\Pran\le6.86$ and $10^{3}\le\Ra^{\prime}\le10^{12}$. 
The standard $k$--$\omega$ model with $C_{\omega b}=1$ yields $T_{\mathit{max}}^{\ast}=0.520 (\Ra^{\prime})^{-0.288}$ for $10^{7}\le\Ra^{\prime}\le10^{10}$, whereas the corrected $k$--$\omega$ model yields $T_{\mathit{max}}^{\ast}=0.325 (\Ra^{\prime})^{-0.245}$.

The modeled scaling behavior of $T_{\mathit{max}}^{\ast}$ in internally heated convection can be interpreted in relation to Rayleigh--B\'enard convection through dimensional analysis. 
Because $T_{\mathit{max}}^{\ast}$ represents temperature normalized by a characteristic heat--flux scale, it scales as $T_{\mathit{max}}^{\ast}\sim\Nu^{-1}$. 
In addition, if the Rayleigh number is estimated as $\Ra \sim g b T_{\mathit{max}}^{\ast} L^3 / (\nu a)$, then $\Ra \sim T_{\mathit{max}}^{\ast}\Ra^{\prime}$ follows.
Substituting these relations into the Rayleigh--B\'enard model scaling $\Nu\sim\Ra^{1/3}$ yields $T_{\mathit{max}}^{\ast}\sim(\Ra^{\prime})^{-1/4}$. 

The key conclusion of this test is that the model constants optimized for Rayleigh--B\'enard convection yield markedly improved predictions even for the qualitatively different problem of internally heated convection.
The widely used choice $C_{\omega b}=1$ strongly underpredicts $\Nu$ in Rayleigh--B\'enard convection (see Figure~\ref{fig:2}b) and correspondingly overpredicts $T_{\mathit{max}}^{\ast}$ in internally heated convection. 
In contrast, the model corrected on the basis of Rayleigh--B\'enard scaling produces substantially improved agreement.

Figure~\ref{fig:10}(b) shows the temperature profiles as a function of $\Ra^{\prime}$, obtained from the corrected model. 
The DNS results of \citet{nourgaliev1997investigation} at $\Ra=1.86\times10^{8}$ exhibit an almost uniform temperature in the bulk region, except near the cooled upper wall. 
The corrected $k$--$\omega$ model reproduces this behavior accurately.

\begin{figure}[h]
	\centering
	\begin{subfigure}{0.48\textwidth}
		\centering
		\includegraphics[scale=1.0]{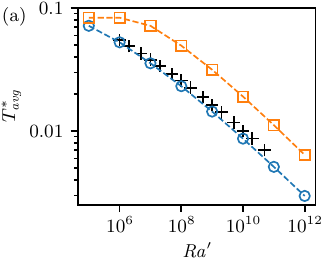}
	\end{subfigure}
	\begin{subfigure}{0.48\textwidth}
		\centering
		\includegraphics[scale=1.0]{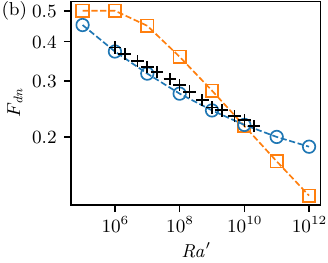}
	\end{subfigure}
	\begin{subfigure}{0.48\textwidth}
		\centering
		\includegraphics[scale=1.0]{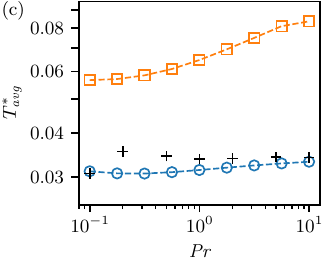}
	\end{subfigure}
	\begin{subfigure}{0.48\textwidth}
		\centering
		\includegraphics[scale=1.0]{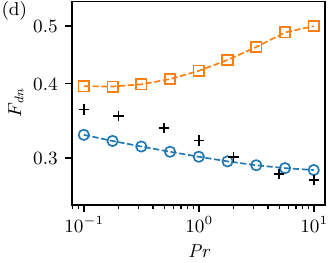}
	\end{subfigure}
	\caption{
		Internally heated convection with top-and-bottom cooling.  
		Circles denote the corrected $k$--$\omega$ model; squares denote the standard $k$--$\omega$ model;  
		plus signs denote the DNS results of \citet{goluskin2016penetrative}.  
		Panels (a) and (c) show the dimensionless average temperature, and panels (b) and (d) show the fraction of heat escaping across the bottom boundary.  
		$\Pran=1$ is used in (a) and (b) to illustrate the dependence on $\Ra^{\prime}$, whereas $\Ra^{\prime}=2\times10^{7}$ is used in (c) and (d) to illustrate the dependence on $\Pran$.
	}
	\label{fig:11}
\end{figure}

Next, Figure~\ref{fig:11} presents the results for the top--and--bottom cooling configuration.  
The parameter ranges are chosen to enable direct comparison with the DNS results of \citet{goluskin2016penetrative}.  
This case contains stably stratified regions with $\mathcal{P}_b<0$ owing to cooling at the lower boundary.  

Figure~\ref{fig:11}(a) clearly shows that the corrected model enhances the overall prediction of $T^{\ast}_{\mathit{avg}}$ over a wide range of $\Ra^{\prime}$.  
As in the top-cooling case, this again demonstrates that the model constants optimized for Rayleigh--B\'enard convection yield markedly improved predictions.

The accuracy of $F_{\mathit{dn}}$ depends on how the model represents stably stratified regions ($\mathcal{P}_b<0$).
As shown in Figures~\ref{fig:11}(b,d), the analytically derived value $C_{\omega b} = -0.5385$ for $\mathcal{P}_b<0$ (\S\ref{sec:4.3}) yields predictions that are noticeably more consistent than those obtained using the conventional choice $C_{\omega b} = -2$ for $\mathcal{P}_b<0$.
The difference arises from how strongly each coefficient damps $k$ under stable stratification.

In Figure~\ref{fig:11}(b), the variation of $F_{\mathit{dn}}$ with increasing $\Ra^{\prime}$ for a given $C_{\omega b}$ is consistent with the mechanism for $\mathcal{P}_b<0$ described in \S\ref{sec:4.3}:
using $C_{\omega b} = -2$ suppresses $k$ more strongly than $C_{\omega b} = -0.5385$, thereby reducing the turbulent heat transport.
A similar behavior occurs near the lower stably stratified region.
Because the uncorrected model employs $C_{\omega b} = -2$, the excessive damping of $k$ in the stable region reduces the turbulent heat transport more rapidly, leading to a steeper decline in $F_{\mathit{dn}}$ as $\Ra^{\prime}$ increases.

Figure~\ref{fig:11}(c) shows that, at fixed $\Ra^{\prime} = 2 \times 10^{7}$, the DNS results yield an approximately constant $T^{\ast}_{\mathit{avg}}$ over $0.1 \le \Pran \le 1$. 
In contrast, the uncorrected model predicts that $T^{\ast}_{\mathit{avg}}$ increases rapidly as $\Pran$ increases. 
This overprediction reflects the same deficiency in the buoyancy formulation that also produces the underprediction of $\Nu$ at large $\Pran$ in Rayleigh--B\'enard convection. 
The relation $T^{\ast}_{\mathit{avg}} \sim \Nu^{-1}$ simply expresses the correspondence between the two quantities, allowing the associated modeling errors to be compared on a common basis.

A similar mechanism underlies the behavior of the downward heat-flux fraction $F_{\mathit{dn}}$ as a function of $\Pran$ in Figure~\ref{fig:11}(d). 
As $\Pran$ increases, the uncorrected model predicts weaker turbulent thermal diffusivity near the upper boundary.
This reduces the heat transported along the upper cold wall and, in turn, causes $F_{\mathit{dn}}$ to rise sharply with $\Pran$, in clear disagreement with DNS. 
The corrected model removes this inconsistency as well, providing a much more accurate prediction of $F_{\mathit{dn}}$ across the entire range of $\Pran$ considered.

\begin{figure}[h]
	\centering
	\begin{subfigure}{0.48\textwidth}
		\centering
		\includegraphics[scale=1.0]{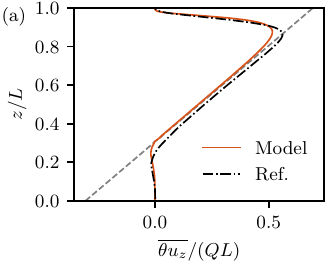}
	\end{subfigure}
	\begin{subfigure}{0.48\textwidth}
		\centering
		\includegraphics[scale=1.0]{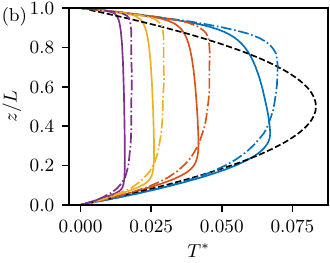}
	\end{subfigure}
	\caption{
		Simulation results for internally heated convection with top-and-bottom cooling using the corrected $k$--$\omega$ model.
		(a) Turbulent heat-flux distribution at $\Ra^{\prime}=10^{7}$ and $\Pran=7$ from the corrected $k$--$\omega$ model (solid line) and the DNS results of \citet{worner1997direct} (dash-dotted line). The dashed line indicates the total heat flux predicted by the model.
		(b) Dimensionless temperature distribution from the corrected $k$--$\omega$ model (solid lines) and the DNS results of \citet{goluskin2016penetrative} (dash-dotted lines). For fixed $\Pran=1$, each line corresponds to $\Ra^{\prime}=10^{6}$, $10^{7}$, $10^{8}$, and $10^{9}$, ordered from right to left. The dashed line indicates the pure-conduction state.
	}
	\label{fig:12}
\end{figure}

Figure~\ref{fig:12} presents the temperature and heat--transport distributions for internally heated convection with top--and--bottom cooling. 
Figure~\ref{fig:12}(a) shows that the corrected $k$--$\omega$ model reproduces the turbulent heat--flux distribution of the DNS of \citet{worner1997direct} with good fidelity. 
In contrast, the temperature profiles in Figure~\ref{fig:12}(b) exhibit clear discrepancies between the model and the DNS. 
The DNS shows an increase of temperature toward the upper boundary, whereas the RANS solution predicts a decrease in temperature in the upper region of the domain.
This discrepancy stems from a basic consequence of the closure formulation. In this configuration, the local heat--flux direction is downward near the cooled lower wall, but upward throughout most of the remaining domain. 
Within the gradient--diffusion framework underlying standard $k$--$\omega$ closures, a positive turbulent heat flux $\overline{\theta u_z}>0$ in the region where $\mathcal{P}_b>0$ necessarily implies that temperature must decrease with height. Consequently, the DNS behavior in the upper region cannot be captured by this class of closures.
Despite this unavoidable inconsistency, the corrected model accurately predicts the heat--flux distribution, the mean temperature, and the partitioning of heat leaving the upper and lower boundaries.
This outcome indicates that the present correction remains sufficiently effective for engineering purposes.

\subsection{Unstably stratified Couette flow}\label{sec:5.2}

To assess the model's applicability beyond purely buoyancy-driven conditions, it is necessary to examine a configuration in which shear and buoyancy interact.  
The unstably stratified Couette flow investigated by \citet{blass2020flow,blass2021effect} provides a suitable test case, representing a mixed-convection regime.

In this configuration, gravity acts in the downward ($-z$) direction and the flow is periodic in the horizontal ($x$) direction.  
The boundary conditions are $U_x = -U_w$ and $T = \Delta$ at $z = 0$, and $U_x = U_w$ and $T = 0$ at $z = L$.  
The Rayleigh and Nusselt numbers follow the definitions used for Rayleigh--B\'enard convection.  
The wall-shear Reynolds number, $\Rey_w = U_w L/\nu$, serves as the control parameter, while the friction Reynolds number, $\Rey_\tau = U_\tau L/\nu$, is treated as an output quantity.

\begin{figure}[h]
	\centering
	\begin{subfigure}{0.48\textwidth}
		\centering
		\includegraphics[scale=1.0]{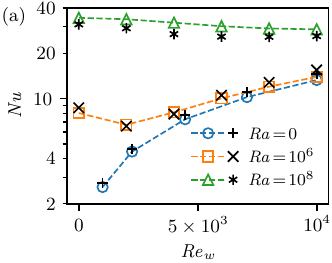}
	\end{subfigure}
	\begin{subfigure}{0.48\textwidth}
		\centering
		\includegraphics[scale=1.0]{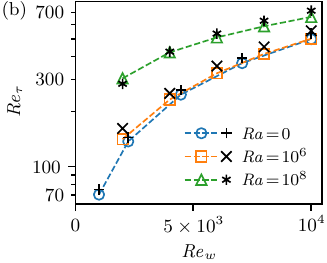}
	\end{subfigure}
	\begin{subfigure}{0.48\textwidth}
		\centering
		\includegraphics[scale=1.0]{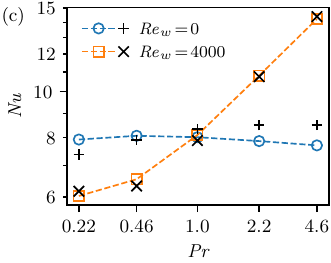}
	\end{subfigure}
	\begin{subfigure}{0.48\textwidth}
		\centering
		\includegraphics[scale=1.0]{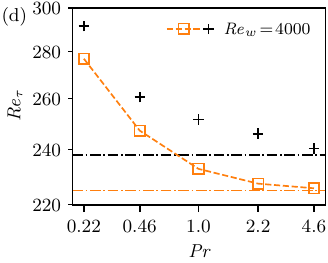}
	\end{subfigure}
	\caption{
		Unstably stratified Couette flow simulation results of the corrected $k$--$\omega$ model. 
		Panels (a,c) show the Nusselt number, and panels (b,d) show the friction Reynolds number, each plotted against one of the two governing parameters:  
		(a,b) variation with the wall--shear Reynolds number at $\Pran = 1$, and  
		(c,d) variation with the Prandtl number at $\Ra = 10^{6}$.  
		Circle, square, and triangle markers denote predictions from the corrected $k$--$\omega$ RANS model.  
		Crosses, plus signs, and asterisks denote DNS data: \citet{yerragolam2022passive} for the $\Ra = 0$ cases in (a,b); \citet{blass2020flow} for the remaining cases in (a,b); and \citet{blass2021effect} for the data in (c,d).  
		The dash--dot horizontal lines in (d) mark the $\Rey_\tau$ values at $\Rey_w = 4000$ and $\Ra = 0$, with the upper line corresponding to the DNS result \citep{blass2020flow} and the lower to the $k$--$\omega$ model.
	}
	\label{fig:13}
\end{figure}

Figure~\ref{fig:13} compares the predictions of the corrected model with DNS results \citep{yerragolam2022passive,blass2020flow,blass2021effect}.  
Panels~(a,b) show the dependence on $\Rey_w$ at $\Pran = 1$, covering a wide range of Rayleigh and Reynolds numbers.  
Across all tested conditions, within $\Ra \le 10^{8}$ and $\Rey_w \le 10^{4}$, the corrected model reproduces both the Nusselt number and the friction Reynolds number with high accuracy.

This behavior can be understood by considering the limiting regimes.  
For $\Ra = 0$, the standard $k$--$\omega$ model predicts the trends of $\Nu$ and $\Rey_\tau$ in passive-scalar Couette flow with errors typically below $10\%$.  
Likewise, in the absence of shear, the present model reproduces the mean Nusselt number of Rayleigh--B\'enard convection to within a similar accuracy.  
DNS results for the mixed regime indicate a smooth transition between these two limits, with no evidence of strong nonlinear interactions between shear and buoyancy for the parameter range considered.  
Consequently, the intermediate behavior is comparatively well captured by the RANS closure.

Panels (c,d) examine the dependence on the Prandtl number by fixing $\Rey_w = 4000$ and $\Ra = 10^{6}$ and varying $0.22 \le \Pran \le 4.6$.
In Figure~\ref{fig:13}(c), the corrected model predicts the mean Nusselt number to within approximately 10\% of DNS.
In Figure~\ref{fig:13}(d), the model consistently underpredicts $\Rey_\tau$ by approximately 5--8\%.
As indicated in the figure, even in the absence of buoyancy, the standard model intrinsically underpredicts $\Rey_\tau$ by about 6\% under comparable conditions.
This baseline discrepancy therefore appears to carry over and yield the overall underprediction in $\Rey_\tau$.
Nevertheless, the proposed buoyancy correction reproduces the increase of $\Rey_\tau$ with decreasing $\Pran$ with high accuracy.

Taken together, these results show that the proposed correction integrates smoothly into the standard $k$--$\omega$ framework and remains robust in a mixed-convection environment where shear and buoyancy interact.

\subsection{Rayleigh--B\'enard convection in a two-dimensional square cavity}\label{sec:5.3}

To evaluate the robustness of the Prandtl-number correction beyond the statistically one-dimensional setting, the model is further assessed in two-dimensional Rayleigh--B\'enard convection within a square cavity. The vertical walls, which are parallel to gravity, are adiabatic, and all boundaries satisfy no-slip conditions.

\begin{figure}
	\centerline{\includegraphics{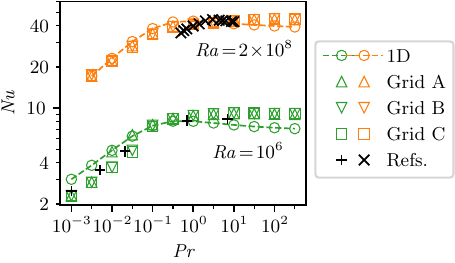}}
	\caption{
		Rayleigh--B\'enard convection simulation results of the corrected $k$--$\omega$ model. 
		Two-dimensional simulations in a square domain are performed using three grid resolutions---Grid~A ($60^{2}$), Grid~B ($100^{2}$), and Grid~C ($150^{2}$)---and compared with the corresponding one-dimensional (1D) simulation.  
		DNS reference data are taken from \citet{pandey2022convective} (plus signs) and \citet{stevens2011prandtl} (crosses).
	}
	\label{fig:14}
\end{figure}

Figure~\ref{fig:14} shows the mean Nusselt number predicted by the two-dimensional simulations across a range of Rayleigh and Prandtl numbers. 
Owing to the presence of large-scale circulation in two dimensions, the heat transfer departs from the one-dimensional result, but the deviation remains moderate. 
For example, the difference between the one- and two-dimensional predictions is within approximately 10\% at $\Ra = 2\times 10^{8}$.
These trends indicate that the Prandtl-number correction formulated under one-dimensional assumptions (\S\ref{sec:4}) retains sufficient accuracy in two dimensions and is capable of reproducing the DNS scaling behavior within the tested parameter space.

Grid independence for the two-dimensional cases is examined using three meshes constructed with a uniform stretching ratio across the entire domain, applied identically in both coordinate directions and symmetric about the domain mid-plane. 
Grid~A employs $60^{2}$ cells with a stretching ratio of 1.20 and a minimum spacing of $1.26 \times 10^{-3}$ (normalized by the domain length) near the walls; 
Grid~B employs $100^{2}$ cells with a stretching ratio of 1.14 and a minimum spacing of $1.19 \times 10^{-4}$; 
Grid~C employs $150^{2}$ cells with a stretching ratio of 1.10 and a minimum spacing of $4.46 \times 10^{-5}$. 

For most Rayleigh--Prandtl combinations, the predicted Nusselt number is effectively grid-independent. 
However, at $\Ra = 10^{6}$ and $\Pran = 10^{-1.5}$ or $10^{-2}$, discrepancies of up to $20\%$ arise among the three meshes. 
Further investigation reveals that this sensitivity originates not from numerical resolution, but from the existence of multiple converged states: even on the same mesh, different initial conditions lead to distinct large-scale circulation patterns.

\begin{figure}
	\centering
	\begin{subfigure}{0.3\textwidth}
		\centering
		\includegraphics[scale=1.0]{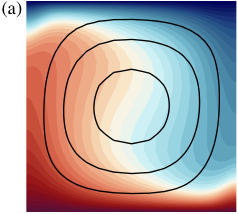}
	\end{subfigure}
	\begin{subfigure}{0.4\textwidth}
		\centering
		\includegraphics[scale=1.0]{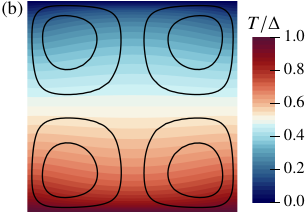}
	\end{subfigure}
	\caption{
		Simulation results from the corrected $k$--$\omega$ model.
		For identical control parameters, $\Ra = 10^{6}$ and $\Pran = 10^{-2}$, two distinct converged states are obtained:
		(a) a single convection cell and (b) four convection cells.
		Black solid lines denote streamfunction contours.
	}
	\label{fig:15}
\end{figure}

Figure~\ref{fig:15} illustrates the two representative steady states observed at $\Pran = 10^{-2}$ and $\Ra = 10^{6}$.
Depending on the initial condition, the solution may converge either to a single-cell circulation with $\Nu = 3.71$, or to a four-cell configuration with $\Nu = 4.81$.
If the large-scale circulation pattern is the same, the predicted Nusselt number is grid-independent across the three meshes.
In contrast, at higher $\Ra$ or $\Pran$, such pronounced changes in the predicted heat flux due to different convection-cell patterns are not observed.

In summary, the presence of large-scale circulation in two dimensions introduces moderate deviations from the one-dimensional predictions. 
However, these effects are typically limited to within $10\%$, and the corrected model maintains its ability to reproduce the proper global $\Nu$--$\Ra$--$\Pran$ trends.
Taken together, these results suggest that the Prandtl-number correction can retain its robustness beyond the one-dimensional setting, supporting its applicability in more complex configurations.

\subsection{Side-heated natural convection in rectangular cavities with differing aspect ratios}\label{sec:5.4}

Natural convection driven by differentially heated vertical walls in an enclosed cavity is one of the most common benchmark problems in prior RANS studies.  
Unlike the earlier cases in \S\ref{sec:5.1}--\ref{sec:5.3}, where buoyancy production dominates the turbulence dynamics, the side-heated cavity is buoyancy-driven but not dominated by buoyancy production.  
Its long use as a validation case, combined with its distinct physical character, makes it a useful additional test for evaluating the robustness of the proposed correction.

The model performance is examined against the benchmark DNS datasets of \citet{trias2007direct,trias2010direct} and \citet{sebilleau2018direct}.  
The computational domain is two-dimensional, defined by $0 \le x \le L$ and $0 \le z \le H$, with gravity acting in the $-z$ direction.  
Two aspect ratios, $H/L = 1$ and $H/L = 4$, are considered.  
The vertical walls are isothermal, with $T=0$ at $x=0$ and $T=\Delta$ at $x=L$, while the horizontal walls are adiabatic.
All boundaries satisfy the no-slip condition.  
The Prandtl number is fixed at $\Pran = 0.71$, and the height-based Rayleigh number, $\Ra_H = g \beta \Delta H^{3} / (\nu \alpha)$, is varied to characterize the resulting heat transfer.

For each pair of $(\Ra_H, H/L)$, simulations are carried out on two grid resolutions and with both the corrected and uncorrected model formulations.  
The meshes employ a uniform stretching ratio across the entire domain and are constructed symmetrically about the mid-plane, with the same number of cells in the $x$ and $z$ directions. 
When expressed in the normalized coordinates $x/L$ and $z/H$, the same grid resolutions as in Section~\ref{sec:5.3} are employed --- Grid B ($100^{2}$) and Grid C ($150^{2}$).  

\begin{figure}
	\centerline{\includegraphics{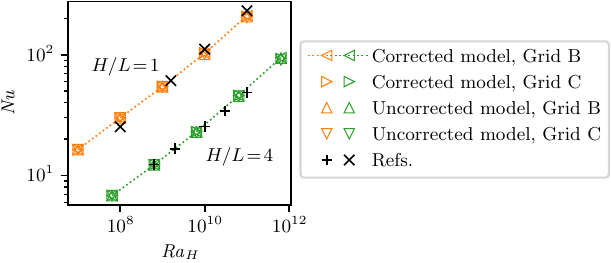}}
	\caption{
		Natural convection in a rectangular cavity with differentially heated vertical walls.
		The overall Nusselt number is shown as a function of the height-based Rayleigh number at $\Pran = 0.71$.
		The upper and lower dotted curves (with adjacent markers) correspond to aspect ratios $H/L = 1$ and $H/L = 4$.
		Four simulation cases are presented using the corrected and uncorrected models on two grids --- Grid~B ($100^{2}$) and Grid~C ($150^{2}$).
		Plus signs and crosses denote DNS data from \citet{trias2007direct,trias2010direct} and \citet{sebilleau2018direct}, respectively.
	}
	\label{fig:16}
\end{figure}

Figure~\ref{fig:16} presents the overall Nusselt number, defined as 
$\Nu = q L / (\alpha \Delta)$, where $q$ is the average heat flux through the vertical walls, as a function of $\Ra_H$ for the two aspect ratios.
For evaluating the uncorrected standard model, the conventional choices $C_{\omega b}=1$ for $\mathcal{P}_b>0$ and $C_{\omega b}=-2$ for $\mathcal{P}_b<0$ are adopted.
Across all four simulation cases, the differences in the predicted Nusselt numbers remain below $1.8\%$, and the mean discrepancy is less than $0.5\%$.  
These results indicate that the solutions are effectively grid-independent and that the buoyancy correction yields a negligible change.

The weak influence of the buoyancy correction follows from the characteristics of this configuration.  
Most of the turbulent heat transport occurs horizontally while gravity acts vertically, so the magnitude of the buoyancy-production term remains small.  
In addition, the upper region of the cavity is warm and the lower region is cold, resulting in downward motion of cooled fluid and upward motion of heated fluid, which yields a negative buoyancy-production term.  
For all cases considered, the integrated negative contribution remains below roughly $5\%$ of the shear production of $k$, indicating that buoyancy plays only a minor role in the modeled turbulence.
This behavior is consistent with the results reported by \citet{peng1999computation}, who examined the same side-heated cavity configuration and reported that predictions using the low-Reynolds-number $k$--$\omega$ model of \citet{wilcox1994simulation} are insensitive to the buoyancy coefficient $C_{\omega b}$ for $1 \le H/L \le 5$ and $10^{10} \le \Ra_H \le 10^{12}$.

\begin{figure}
	\centering
	\begin{subfigure}{0.25\textwidth}
		\centering
		\includegraphics[scale=1.0]{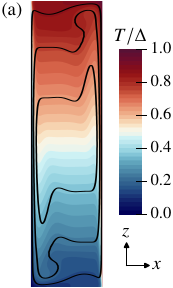}
	\end{subfigure}
	\begin{subfigure}{0.35\textwidth}
		\centering
		\includegraphics[scale=1.0]{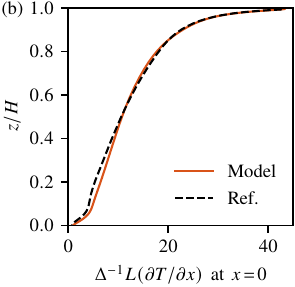}
	\end{subfigure}
	\caption{
		Simulation results of vertically heated natural convection for $H/L = 4$ and $\Ra_H = 6.4\times10^{8}$:
		(a) temperature field with streamfunction contours (black solid lines);
		(b) local heat-flux distribution along the cold wall (solid line).
		The dashed line denotes the DNS results of \citet{trias2007direct}.
	}
	\label{fig:17}
\end{figure}

Further detail is shown in Figure~\ref{fig:17} for the case $H/L = 4$ and $\Ra_H = 6.4 \times 10^{8}$.  
Figure~\ref{fig:17}(a) shows the temperature field and the streamfunction contours predicted by the model, which reproduce the expected features of side-heated natural convection, including thin thermal boundary layers along the isothermal vertical walls and corner-induced recirculation cells.  
Figure~\ref{fig:17}(b) presents the corresponding local heat-flux distribution along the cold wall.  
Here, the results are obtained using the corrected model, though the uncorrected formulation produces nearly identical profiles.  
The mean value of the normalized heat-flux profile in Figure~\ref{fig:17}(b) corresponds to an overall Nusselt number of 12.3.
Comparison with the DNS data of \citet{trias2007direct} shows that the standard $k$--$\omega$ model reproduces the heat-transfer distribution with high accuracy.

In summary, the side-heated cavity provides a useful check that the proposed buoyancy correction does not degrade the performance of the standard model in a shear-dominated regime.  
For this configuration, the standard $k$--$\omega$ closure already reproduces the global Nusselt number and the local heat flux distribution with high accuracy, and the corrected and uncorrected formulations yield virtually identical results on two grid resolutions.  
Combined with the buoyancy-dominated cases examined in \S\ref{sec:5.1}--\ref{sec:5.3}, these observations indicate that the proposed correction enhances the model response when buoyancy production is significant, while having negligible influence in configurations where buoyancy plays only a limited role.

\section{Discussion}\label{sec:6}

To avoid the possibility that differences in modeling philosophy could be used to justify an unbalanced or dismissive evaluation of this manuscript, the following discussion clarifies the conceptual stance underlying the present work. 
This perspective closely aligns with that articulated in Wilcox's textbook \citep{wilcox2006turbulence} and in Spalart's review on turbulence-modeling philosophies \citep{spalart2015philosophies}. 
However, because these references offer only limited treatment of buoyancy effects, this section focuses specifically on organizing and elaborating the modeling rationale for buoyancy-driven turbulence.

\subsection{On the interpretation of the present modeling approach}

Critiques that the present approach is empirical are consistent with the inherent nature of RANS modeling. 
The primary objective of RANS is not to uncover new fundamental turbulence physics, but rather to construct computationally tractable closures that reproduce observed statistical behavior with reasonable fidelity \citep{pope2000turbulent,wilcox2006turbulence}.  

Two-equation RANS models do not contain any hidden or mysterious physical meaning. 
Instead, they are built upon mathematical structures designed to remain broadly consistent with canonical turbulent behaviors. 
For example, the standard $k$--$\omega$ model accommodates the logarithmic velocity law with a von K\'arm\'an constant, boundary-layer scaling, spreading rates of free-shear flows, adverse-pressure-gradient effects, and self-similarity in homogeneous shear and decaying turbulence \citep{wilcox2006turbulence}. 
When a model performs well across substantially different canonical problems, a degree of robustness is often observed when it is applied to mixed configurations that combine multiple flow mechanisms. 
At the level of Reynolds-averaged quantities, and for the purpose of predicting trends in dimensionless outputs, strongly nonlinear cross-coupling effects arising solely from the coexistence of two distinct flow mechanisms are relatively uncommon.  

From this perspective, the central working hypothesis underlying the present model can be articulated as follows:  
a closure that can reasonably predict the Nusselt number for Rayleigh--B\'enard convection across a broad $\Ra$--$\Pran$ parameter space is likely to possess meaningful predictive capability for other buoyancy-dominated flows.  
This hypothesis has been critically examined in the present study by testing the model against a range of configurations. 
The fact that the present model yields markedly improved predictions in these cases lends empirical support to this guiding assumption.  

In a similar vein, RANS models are best regarded as a hypothetical framework that is distinct from the actual dynamical mechanisms operating in real turbulent flows. 
Consequently, directly mapping phenomenological turbulence theories such as the Grossmann--Lohse (GL) framework \citep{grossmann2000scaling} onto RANS equations on a one-to-one basis must be treated with caution. 
This does not imply that phenomenological theories are incorrect; rather, it reflects the fact that real turbulent flows and RANS modeling operate at fundamentally different levels of abstraction.  

Grossmann--Lohse theory \citep{grossmann2000scaling} derives scaling laws based on a large-scale coherent velocity, whereas RANS does not explicitly resolve plume structures or coherent circulation patterns. 
Instead, their net influence is represented through an enhanced turbulent thermal diffusivity distributed across the domain. 
This distribution is governed primarily by the empirical balance between buoyant production and dissipation in the $\omega$ equation, rather than by explicitly resolved turbulent motions.  

In this study, a statistically one-dimensional RANS solution for $U=0$ is first derived, after which its predictions are shown to remain consistent when extended to two-dimensional configurations. 
This indicates that meaningful heat transfer can be captured within a RANS framework even in the absence of a resolved mean circulation. 
This result should be interpreted as a property of the RANS closure framework, rather than as a challenge to phenomenological theories.  

For similar reasons, a term-by-term comparison in which parameters measured directly from DNS are transplanted into a RANS framework is not always appropriate. 
Many second-moment closures, in practice, still rely on a two-equation model as their backbone to close the dissipation, thereby circumventing the formidable difficulties associated with modeling genuinely non-local terms such as pressure--velocity correlations.  

In the turbulent kinetic energy budget, the pressure--velocity correlation term is often relatively small in many shear-driven flows and can be neglected with limited practical consequence \citep{pope2000turbulent}. 
In buoyancy-dominated flows, however, thermal plumes impinging on solid boundaries can generate contributions that are comparable to---or even larger than---the local production terms \citep{togni2015physical}. 
Because RANS closures do not represent this mechanism explicitly, the near-wall distributions of $k$, $\varepsilon$, and $\nu_T$ inevitably deviate from those observed in DNS. 
Under such circumstances, even if quantities such as the turbulent Prandtl number $\Pran_T$ were extracted from DNS and imposed directly in a RANS model, there would be no guarantee that the resulting predictions would be physically consistent or accurate.

Precisely this issue explains why analytical solutions for canonical flows must be derived in order to understand the intrinsic behavior of standard RANS models. 
This viewpoint is repeatedly emphasized in \citet{wilcox2006turbulence}, where it is cautioned that ``an important point to keep in mind is to avoid modeling the differential equations rather than the physics of turbulence.''

RANS has inherent limitations in representing complex large-scale dynamics. 
Nevertheless, it retains substantial practical value. 
If a model can predict Nusselt numbers and temperature distributions within roughly ten percent accuracy at a computational cost orders of magnitude lower than DNS or large eddy simulation, this constitutes a significant engineering achievement. 
Moreover, attaining such performance typically requires a careful understanding of the closure structure rather than simple parameter fitting.

\subsection{On the justification of the level of abstraction in RANS models}

A recurring debate in buoyancy-driven turbulence modeling concerns whether 
eddy-viscosity and gradient-diffusion models are fundamentally inadequate and must therefore be replaced by higher-order closures.
Different modeling philosophies assign different priorities to physical detail, generality, and computational robustness. 
Accordingly, the appropriate level of closure should be regarded as a strategic choice shaped by the intended application rather than a binary distinction between correct and incorrect approaches.

The primary working objective of this study is to reproduce a wide range of buoyancy-driven turbulent flows through minimal modifications to the standard two-equation RANS framework. 
The principal quantitative evaluation criteria are the accurate prediction of mean temperature distributions and turbulent heat transfer, which are of primary interest in engineering applications.  
Within this framework, it is shown that introducing only two algebraic, dimensionless corrections to the standard $k$--$\omega$ model is sufficient to improve its performance across a broad range of buoyancy-driven problems.

Higher-order closures, by contrast, seek to represent individual terms in the turbulence budget equations in a more detailed, term-by-term manner, often guided by DNS data, realizability constraints, and wall-asymptotic arguments. 
Such models can provide deeper physical insight into turbulence anisotropy, but they also entail greater mathematical complexity and more extensive calibration. 

Although it is sometimes argued that higher-order approaches are ``physically justified'' because they are derived from the transport equations of turbulence moments, this does not remove the need for empiricism. 
In practice, most higher-order models still rely on dimensional arguments, on idealizations of isotropic turbulence that neglect non-local effects, and on ad hoc scale-determining equations introduced to close the dissipation. 
Consequently, many higher-order closures continue to depend on empirical modeling assumptions that are not qualitatively different from those used in standard two-equation models, even if they are embedded within a more elaborate mathematical structure.  

For this reason, the claim that higher-order models inherently capture the underlying physics whereas low-order approaches such as the gradient-diffusion hypothesis do not is not compelling. 
The substantive difference lies less in which class of model is more ``physical,'' and more in how empirical assumptions are formulated and constrained within the closure.

\subsection{The practical role of higher-order closures}

Advocates of higher-order turbulence closures often argue that standard low-order models have reached an inherent performance limit and that meaningful further progress requires a transition to a more complex modeling hierarchy. 
This “glass ceiling” view implicitly assumes that prediction errors mainly arise from fundamental flaws in the baseline framework itself, rather than from unresolved issues within it. 
However, at least for buoyancy-driven flows, this premise is difficult to sustain. 
The present study shows that the predictive capability for buoyancy-driven turbulence can be substantially improved within the existing two-equation framework using only minimal algebraic modifications.

Algebraic heat flux models (AFM), or more generally second-moment closures, are frequently presented as advanced alternatives to the gradient-diffusion hypothesis (GDH) \citep{hanjalic2002one}. 
AFM is obtained by applying a weak-equilibrium assumption to the turbulent heat-flux transport equation—thereby neglecting material-derivative terms—and by decomposing the fluctuating pressure-related terms into slow and rapid components following \citet{lumley1979computational}. 
Nevertheless, in most buoyancy-driven flows, the dominant contribution to the turbulent heat flux remains the mean temperature gradient term, which effectively reduces the closure to a GDH-like form.

Consequently, replacing GDH with AFM typically leads to only minor changes in the predicted Nusselt number \citep{hadvziabdic2025revisiting,joo2025global}, often on the order of 4--6\%. 
Buoyancy-related higher-order terms can be amplified to improve agreement for specific cases, but this strategy risks overfitting and may compromise general applicability. 
From a practical standpoint, higher-order closures are therefore best regarded as refinements that become useful only after a reliable low-order baseline has been established, rather than as wholesale replacements for standard two-equation models.

As discussed in \S\ref{sec:2.2}, a more fundamental issue in buoyancy-RANS modeling is that the baseline structure of standard two-equation models has not been sufficiently examined for thermal convection problems. 
For this reason, priority is given to a minimal, algebraic modification of the standard $k$--$\omega$ framework instead of immediately adopting a more complex closure hierarchy. 
This strategy does not preclude the future incorporation of higher-order terms; rather, it provides a consistent and transparent starting point for such developments.

Ultimately, model validity should be assessed primarily by robust predictive performance across diverse flow regimes, particularly with respect to mean Nusselt numbers and temperature distributions. 
Higher-order models may better represent turbulence anisotropy, but if they perform well only in narrowly defined cases while failing to predict heat-transfer trends systematically over broad parameter ranges, their practical value remains limited.

\subsection{Accuracy versus simplicity}

The primary motivation for deriving an analytical solution is to clarify the underlying behavior of the model. To obtain an explicit form, a degree of simplification is unavoidable. Consequently, some level of quantitative discrepancy must be accepted. Nevertheless, the derivation is justified because it captures the dominant trends across the relevant parameter space with favorable fidelity.

It is acknowledged that the $\Ra$--$\Pran$ dependence of the Nusselt number may vary across different regions of parameter space. Nevertheless, regime-dependent $\Pran$ scaling is deliberately not imposed as an explicit function of $\Ra$.

Although Grossmann--Lohse theory \citep{grossmann2000scaling} suggests regime-dependent behavior, the available datasets are not yet sufficiently comprehensive to support robust and defensible model calibration across all regimes. Parameter fitting is therefore based on recent high-fidelity DNS data within the $\Ra$--$\Pran$ range corresponding to Fig.~\ref{fig:5}(b). Within this domain, there is no clear evidence of sharp, systematic scaling transitions that would justify additional model complexity. While systematic deviations of the $\Nu\sim\Ra^{m}$ exponent $m$ for $\Pran<1$ and $\Ra>10^8$ are plausible, existing data do not yet provide a sufficient basis for introducing further corrections.

Moreover, the primary objective of a standard RANS model is not to reproduce asymptotic scaling exponents exactly, but to predict mean heat transfer within an acceptable error margin across a broad range of conditions. From this standpoint, the present level of deviation from $m=1/3$ is tolerable. Additional refinements could improve local accuracy, but would likely offer limited practical benefit relative to the increase in model complexity.

The transition to an ultimate regime with $m>1/3$ remains an active topic of debate \citep{lohse2023ultimate}; however, experimental data beyond $\Ra \ge 10^{11}$ are currently too sparse for reliable model calibration. Accordingly, the intended applicability of the present model is restricted to $\Ra< 10^{11}$. A quantitative treatment of the ultimate regime within a RANS framework will require further study of the interaction between large-scale circulation and turbulent boundary layers near the walls.

There are, of course, many additional physical considerations that are not addressed here, including wall-limit behavior for each variable, realizability constraints, and anisotropy effects. It is even questionable whether it is desirable for a RANS model to reproduce all of these features exactly in every detail. Nevertheless, if an individual researcher judges that the accurate representation of a particular physical quantity is critically important, further model refinements can in principle be introduced for that specific purpose.

Moreover, work of this kind must fully account for the nonlinear interactions among the model terms of the type described in Eq.~(\ref{eq: aT wall restriction}). When this requirement is overlooked, instability issues are frequently encountered. These challenges are substantially more difficult to diagnose and manage in higher-order closures because of their increased model complexity.

Consistent with the overall philosophy of this paper, this study does not view the continual addition of complexity to turbulence models in this manner as a desirable path forward. There is no “free lunch”: from a practical engineering perspective, pursuing marginal performance gains by substantially increasing model complexity is considered an unfavorable trade-off.  

Overall, RANS development inevitably involves a balance between simplicity and accuracy. By preserving the core structure of the Wilcox model and introducing only minimal corrections, generality and numerical robustness are retained while substantially improved predictive performance for buoyancy-driven flows is achieved.

\section{Conclusions}\label{sec:7}

This study establishes an analytical framework for representing buoyancy effects in the standard $k$--$\omega$ model, addressing the long-standing lack of a consistent formulation for two-equation closures. 

To that end, an explicit analytical solution of the $k$--$\omega$ model in statistically one-dimensional Rayleigh--B\'enard convection is derived.
This solution reveals how the conventional buoyancy modeling determines the heat-transport scaling in natural convection and clarifies the origin of the model’s deviations from established $\Nu$--$\Ra$--$\Pran$ trends.
Guided by this analysis, the closed-form correction functions emerge naturally rather than from empirical tuning, allowing the model to recover the observed scaling behavior. 
Only two algebraic, dimensionless functions are introduced, and the formulation reduces to the original model in the absence of buoyancy, ensuring seamless compatibility and negligible computational overhead. 
The associated modeling coefficients are calibrated through a transparent and reproducible process using literature DNS and experimental datasets.

The corrected model is evaluated across a wide range of buoyancy-driven configurations---including Rayleigh--B\'enard convection in one- and two-dimensional settings, two internally heated convection systems, unstably stratified Couette flow, and vertically heated natural convection with varying aspect ratios. 
Across all cases, the corrected formulation substantially improves quantitative accuracy relative to the baseline model while maintaining robustness and numerical stability.

In summary, the proposed framework provides a systematic treatment of buoyancy effects within two-equation RANS closures and yields consistently accurate predictions of turbulent heat transport across a broad range of buoyancy-driven configurations and parameter regimes.

Despite these strengths, several limitations remain outside the present scope. 
Although the model reproduces the turbulent heat flux distribution accurately, its reliance on the gradient--diffusion hypothesis means that the temperature field may exhibit modest discrepancies in situations where the turbulent heat flux is not well aligned with the mean temperature gradient.
The calibration dataset could also be expanded, particularly at high $\Ra$ and low $\Pran$, where reference data remain scarce.  
Finally, other physical effects that can alter natural-convection dynamics, such as system rotation, are not considered here.
These limitations indicate some directions for further refinement and for extending the present approach to more complex buoyancy-affected flows.

\section*{ACKNOWLEDGMENTS}
This work did not receive any specific grant from funding agencies in the public, commercial, or not-for-profit sectors.

\section*{Conflict of Interest}
The author has no conflicts to disclose.

\section*{Data availability statement}
Data sharing is not applicable to this article as no new data were created or analyzed in this study.

	\bibliography{aipsamp}
		
	\end{document}